\documentclass[epj,nopacs]{svjour}
\usepackage{graphics}

\usepackage{hyperref}
\usepackage{graphicx}%
\usepackage{multirow}%
\usepackage{amsmath,amssymb,amsfonts}%
\usepackage{mathrsfs}%
\usepackage{xcolor}%
\usepackage{textcomp}%
\usepackage{manyfoot}%
\usepackage{booktabs}%
\usepackage{algorithm}%
\usepackage{algorithmicx}%
\usepackage{algpseudocode}%
\usepackage{listings}%
\usepackage{physics}
\usepackage{upgreek}
\usepackage{adjustbox}
\usepackage{subfig}
\usepackage{gensymb}

\graphicspath{{figs/}}

\newcommand{\gag}{g_{a\gamma}}
\newcommand{\Pgg}{P_{\gamma\gamma}}
\newcommand{\bigO}[1]{\mathcal{O}(#1)}
\DeclareMathOperator{\diag}{diag}

\begin{document}
\title{Projected sensitivity of CTAO to axion-like particles from blazars with a machine learning approach}
\author{F. Schiavone\inst{1,2}\thanks{e-mail: francesco.schiavone@ba.infn.it} \and L. Di Venere\inst{2} \and F. Giordano\inst{1,2}
%
}                     
%
%
\institute{Dipartimento Interateneo di Fisica ``Michelangelo Merlin'', Università degli Studi di Bari, Via Amendola 173, 70125, Bari, Italy
\and INFN Sezione di Bari, Via Amendola 173, 70125, Bari, Italy}
\date{Received: date / Revised version: date}
%
\authorrunning{F. Schiavone \and L. Di Venere \and F. Giordano}
\titlerunning{CTAO sensitivity to axion-like particles from blazars}

\abstract{
Blazars are a class of active galactic nuclei, supermassive black holes located at the centres of distant galaxies characterised by strong emission across the entire electromagnetic spectrum, from radio waves to gamma rays. Their relativistic jets, closely aligned to the line of sight from Earth, are a rich and complex environment, characterised by the presence of strong magnetic fields over parsec-scale lengths. Owing to their cosmological distance from Earth, these sources serve as ideal targets to probe non-standard gamma-ray propagation. In particular, axion-like particles (ALPs) could be detected through their coupling to photons, which enables ALP-photon conversions in external magnetic fields, leading to distinct signatures in the blazars’ gamma-ray spectra. In this work, we estimate the potential of the Cherenkov Telescope Array Observatory (CTAO) to constrain the ALP parameter space by simulating observations of two bright blazars, Mrk 501 and PKS 2155$-$304. We obtain projected $2\sigma$ exclusion regions, demonstrating that CTAO will be able to consistently improve present limits thanks to its greater energy resolution and point-source sensitivity with respect to present ground-based gamma-ray telescopes. In addition to the standard statistical technique based on the likelihood ratio test, we further demonstrate the application of a new method based on machine learning classifiers, which may help in reducing the effect of systematic model-dependent uncertainties in future ALP searches.
} 
\maketitle
\section{Introduction}
\label{sec:intro}

Axion-like particles (ALPs) are a kind of hypothetical, pseudo-scalar elementary particles whose existence is predicted by several extensions of the Standard Model. Their name is derived from the original axion model, which was formulated by Peccei and Quinn \cite{pecceiquinn1977}, Weinberg \cite{weinberg1978}, and Wilczek \cite{wilczek1978} as a solution of the strong CP problem in the context of quantum chromodynamics (QCD). While the existence of this original axion was soon ruled out \cite{donnelly1978,zehnder1981}, several ``invisible axion'' models were proposed, such as KSVZ \cite{kim1979,shifman1980} and DFSZ \cite{dine1981,zhitnitsky1980}, which could elude experimental constraints while still solving the strong CP problem. In addition to these, it became clear that axion-like particles could be a more general prediction of beyond-the-Standard-Model theories such as Grand Unification Theories and string theories. These ALPs are not necessarily related to the original Peccei-Quinn model, but, like the QCD axion, they often arise as pseudo-Nambu-Goldstone bosons of spontaneously broken symmetry groups. Axions and ALPs are currently of great interest in cosmology and astrophysics, as they are among the best-motivated candidates as constituents of dark matter. Reviews of models and experimental searches for both QCD axions and ALPs can be found e.g. in refs. \cite{jaeckel2010,ringwald2012,irastorza2018,diluzio2020}.

The most important feature for the experimental detection of ALPs is their effective two-vertex coupling to photons, whose strength is controlled by the coupling constant $\gag$. Such a coupling is the basis for the most common approach to ALP searches, which exploits the expected ALP-photon mixing in the presence of external magnetic fields \cite{sikivie1983,raffelt1988}. This effect is employed, for example, in beam regeneration (or ``light shining through walls'', LSW) experiments, such as ALPS-II \cite{bahre2013}, where a laser beam is shot against an opaque screen; in the presence of a strong enough magnetic field, some photons in the beam should be able to convert into ALPs, pass through the screen and oscillate back to photons on the other side. Other experiments for the direct detection of ALPs include helioscopes, like CAST \cite{zioutas1999} and the foreseen IAXO \cite{armengaud2019}, and resonant cavity experiments such as ADMX \cite{du2018} and MADMAX \cite{brun2019}.

As a general property of QCD axions, the axion-photon coupling $\gag$ is proportional to the axion mass $m_a$, as both quantities are inversely proportional to the Peccei-Quinn energy scale $f_a$ (e.g. \cite{irastorza2018}). On the other hand, in the more general case of ALPs there is typically no fixed relationship between $m_a$ and $\gag$, so that they are independent coordinates spanning a 2-dimensional parameter space. 

Astrophysical gamma-ray sources characterised by one or more magnetic field environments along their line of sight to Earth offer a natural framework to probe ALP-related effects, which has been extensively discussed in literature \cite{hooper2007,mirizzi2007,mirizzi2008,deangelis2007,deangelis2008,deangelis2011,horns2012,batkovic2021,galanti2022}. Gamma-ray polarization effects could in principle observed, as ALPs only couple to the photon state parallel to the external transverse magnetic field; however, such effects are not detectable with present-day instruments \cite{bassan2010,galanti2024}. On the other hand, two main measurable spectral features may be investigated, whose extent generally depends on the field strength and topology, ALP parameters, source distance, and energy range of interest. The first is the presence of energy-dependent \textit{wiggles} superimposed on the spectral shape, due to the peculiar shape of the photon survival probability $\Pgg$ as a function of energy. The second is a spectral \textit{hardening} at very high energy (VHE), due to gamma-ray photons converting into ALPs in the magnetic field of the source and back into photons in the Milky Way. This mechanism would allow the ALP-photon beam to travel cosmological distances, eluding the absorption that is normally experienced by VHE gamma rays due to the pair production (Breit-Wheeler scattering \cite{breitwheeler1934}) on the soft photons making up the extragalactic background light (EBL) \cite{dominguez2011,franceschini2008,finke2010,gilmore2012,saldanalopez2021,deangelis2013}. The resulting effect, analogous to a ``cosmic LSW experiment'', is expected to be more significant in the spectral tails of distant sources, for which the EBL absorption in the absence of ALPs would be strongest.

A typical target of these studies is represented by blazars. These sources are a particular class of active galactic nuclei (AGN), supermassive black holes characterised by accretion disks and strong relativistic jets at a small angle ($\lesssim 20\degree$) with the line of sight from Earth \cite{padovani2017}. Their broadband spectral energy distribution (SED) is characterised by 
a low-energy component due to the synchrotron radiation emitted by accelerated electrons in the jet, peaked between infrared and X-ray frequencies, and higher-energy emission in the gamma rays that may be produced either by leptonic or hadronic mechanisms \cite{bottcher2013,cerruti2020}.
Leptonic mechanisms typically assume the gamma-ray emission to be originated by the same electron population that produces the synchrotron radiation via inverse Compton scattering, either on the synchrotron radiation field itself (synchrotron self-Compton, SSC) or on an external low-energy photon field (such as thermal photons from the broad line region, accretion disk or dusty torus, or the cosmic microwave background). In hadronic models, high-energy photons are emitted by a population of relativistic protons and other nuclei in the jet, typically via synchrotron radiation or photo-meson production, with the associated production of neutrinos.
Based on their optical properties, blazars can be classified between BL Lacertae (BL Lac) objects and flat spectrum radio quasars (FSRQs). In addition, they can be distinguished between \mbox{high-,} intermediate- and low-synchrotron peaked (HSP, ISP and LSP) depending on their peak synchrotron frequency. Many blazars display significant spectral variability over timescales ranging from years down to minutes \cite{bottcher2019}, with high-flux flaring states representing important opportunities of detection or even discovery of new sources. Despite being a very rare class of objects, almost all extragalactic sources detected in gamma-rays are associated with blazars. 

The relativistic jets of blazars typically feature magnetic fields of order $\bigO{\rm mG}$ to $\bigO{\rm G}$ over $\bigO{\rm pc}$ scales. These objects may also be located inside galactic clusters with complex magnetic field structures of $\bigO{\rm\upmu G}$ intensity over $\bigO{\rm Mpc}$ scales; a notable example of this environment is represented by the Perseus galaxy cluster, which hosts the Seyfert galaxy NGC~1275. Additionally, being extragalactic sources, blazars are located at a cosmological distance from Earth (represented by the redshift parameter $z$), which makes EBL absorption effects important. All these characteristics may allow for hints of ALP-photon conversions in the gamma-ray spectra of blazars. Several constraints on the ALP parameter space have been obtained by analysing gamma-ray data from different blazars \cite{abramowski2013,jacobsen2023,davies2023} and from the Perseus galaxy cluster \cite{ajello2016,abe2024}.

In this context, the next-generation Cherenkov Telescope Array Observatory (CTAO\footnote{\href{https://www.ctao.org/}{https://www.ctao.org/}}, formerly known as CTA) will be an ideal instrument to probe this kind of phenomena. The observatory will be made up of three different types of imaging atmospheric Cherenkov telescopes (IACTs), distributed in arrays over two sites in the Northern and Southern hemispheres \cite{acharya2013}. The Northern Array will be located at Observatorio del Roque de los Muchachos in La Palma, Canary Islands, and will be made up of 4 Large-Sized Telescopes (LSTs) and 9 Medium-Sized Telescopes (MSTs); the Southern Array, in Paranal, Chile, will feature 14 MSTs and 37 Small-Sized Telescopes (SSTs). These different configurations were designed to obtain the best sensitivity to sources visible from the two hemispheres, with the Southern array optimised to detect the highest energy photons (up to several hundreds of TeV) from the Galactic centre. In addition to this baseline ``Alpha'' configuration, an upgrade is foreseen by the Italian CTA+ Program\footnote{\href{https://pnrr.inaf.it/progetto-ctaplus/}{https://pnrr.inaf.it/progetto-ctaplus/}} to add 2 LSTs and 5 more SSTs to the Southern site \cite{loporchio2025}.  
An alternative configuration has also been proposed, using telescopes of the Schwarzschild-Couder type \cite{adams2021} to complement the standard MSTs (which feature a Davies-Cotton optical design). CTAO will obtain a significant improvement in all areas of performance with respect to present-days IACT systems: in particular, its greatly enhanced energy resolution and point-source sensitivity will make it an ideal instrument to study the spectra of distant sources in search of potential non-standard spectral features \cite{ctascience2018}. The CTAO sensitivity over the ALP parameter space has been estimated by simulating observations of various blazars \cite{meyer2014a,meyer2014b} and of the Perseus galaxy cluster \cite{abdalla2021}.

Almost all of the previously cited works rely on the definition of some kind of test statistic (TS) based on the logarithm of the likelihood ratio (equivalently, the difference in log-likelihoods) to test the ALP hypothesis on real or simulated data \cite{jacobsen2023,davies2023,ajello2016,abe2024,meyer2014a,meyer2014b,abdalla2021}. In mathematical terms, the TS is given by
\begin{equation}
    {\rm TS} = -2\ln\frac{\mathcal{L}(H_1)}{\mathcal{L}(H_0)} \,,
\end{equation}
where $H_i$ denotes a generic hypothesis and $\mathcal{L}$ is the likelihood function. The latter is defined as the probability that the model parameters take on certain values under $H_i$, conditioned to the observed data. In this work, we introduce a different approach for constraining the ALP parameter space with binned gamma-ray data, based on the use of machine learning (ML) classifiers. We apply our method to simulated CTAO observations of selected blazars, and compare the results with those obtained by using the standard likelihood-ratio test (LRT).

This paper is structured as follows. In Section \ref{sec:sources} we describe the blazar selection, spectral modelling and simulation, as well as the modelling of ALP-photon oscillations. In Section \ref{sec:lr-limits} we review the standard LRT method to derive upper limits on the ALP parameter space, while in Section \ref{sec:ml-limits} we describe an alternative approach to the same problem based on the use of ML classifiers. In Section \ref{sec:discussion} we discuss and compare the results obtained with the two methods for the selected sources, and in Section \ref{sec:conclusions} we present our conclusions.

\section{Source modelling and simulation}
\label{sec:sources}
\subsection{Source selection}
The observation of AGNs is one of the Key Science Projects (KSPs) proposed for CTAO, to be carried out within the first 10 years of operation of the observatory \cite{ctascience2018}. The programme of this KSP will be organized in three parts: long-term monitoring of known AGNs, search and follow-up of flares, and measurement of high-quality spectra for different AGN classes and redshifts. One of the declared goals of the AGN KSP is to study possible imprints of new fundamental physics on the cosmological propagation of gamma-rays, including ALP searches \cite{ctascience2018,abdalla2021}.

In this work we shall focus on two of the most well-known and studied TeV blazars: Markarian 501 (Mrk~501) and PKS~2155$-$304. Both are high-synchrotron peaked BL Lac objects (HBLs), featuring significant emission at TeV-scale energies, and are included in the plans for long-term monitoring and high-quality spectral measurements of the KSP described above. This guarantees that CTAO will be able to gather a large amount of good quality data from these sources, which will be available for applying our proposed analysis.

The VHE spectral energy distribution (SED) of blazars is usually modelled with a simple function of energy, such as a power law (PL). Other common models include a log-parabola (LP)
\begin{equation}
\label{eq:lp}
    \phi(E) = \dv{N}{E}=\phi_0\qty(\frac{E}{E_0})^{-\alpha-\beta\log(E/E_0)}\,,
\end{equation}
which reduces to a simple power-law for $\beta\to0$, 
or a power law with exponential cutoff (ECPL)
\begin{equation}
\label{eq:ecpl}
        \phi(E) = \dv{N}{E}=\phi_0\qty(\frac{E}{E_0})^{-\alpha}e^{-E/E_{\rm cut}}\,.
\end{equation}
 For the baseline states of our sources, we adopt the best-fit spectral models reported in the 4th \textit{Fermi}-LAT Source Catalog Data Release 4 (4FGL-DR4) \cite{4fgl-dr3,4fgl-dr4}. For Mrk~501 and PKS~2155$-$304, these are simple log-parabola functions. Since there are no observational constraints on the gamma-ray emission of blazars above a few tens of TeV, we conservatively assume that the validity of these models can be extended up to $10\,{\rm TeV}$ (thus not fully exploiting the CTAO energy range). In other works, an exponential cutoff factor around $1\,{\rm TeV}$ in rest-frame energy was added to these spectra (e.g. \cite{abdalla2021}). We have checked that this modification does not significantly impact the extent of our final results.

As mentioned in the introduction, blazars are extragalactic objects, located at a distance measured by the cosmological redshift parameter $z$. For such targets, the absorption of gamma rays due to scattering on EBL photons must be taken into account. This effect is commonly quantified by an optical depth $\tau(E;z)$, such that the absorbed gamma-ray flux is attenuated by an exponential factor $\exp[-\tau(E;z)]$. Then, in the standard astrophysical scenario, the observed gamma-ray flux at Earth $\phi_{\rm obs}$ is related to the intrinsic gamma-ray flux at the source $\phi_{\rm int}$ by
\begin{equation}
    \phi_{\rm obs}(E)=\phi_{\rm int}(E)e^{-\tau(E;z)}\,.
\end{equation}
Since the effect of EBL absorption is negligible in the \textit{Fermi}-LAT energy range, we take the 4FGL models as representative of the intrinsic spectra of the sources, and apply the EBL absorption factor to account for flux reduction at higher energies. Several models of the EBL exist in the literature, both phenomenological and obtained from first principles, mostly providing consistent results \cite{dominguez2011,franceschini2008,finke2010,gilmore2012,saldanalopez2021,deangelis2013}. In the following, we shall employ tabulated values of the optical depth as computed in the model of Dom\'inguez et al. \cite{dominguez2011}. We discuss the impact of choosing different EBL models on our final results in Appendix \ref{app:systematics}.

In addition to the \textit{Fermi}-LAT baseline states, we consider for each source a flaring state taken from previous VHE measurements. Specifically, we take the Mrk~501 flare observed by the MAGIC-I telescope on 30 June 2005 \cite{albert2007}, which is again well-fit by a LP shape, and the historical PKS~2155$-$304 flare observed by H.E.S.S. on 29 July 2006 \cite{aharonian2009}, which follows an ECPL. These spectra are corrected for EBL absorption and, as in the previous case, we assume their validity up to $10\,{\rm TeV}$ and we extrapolate them down to $30\,{\rm GeV}$.

The selection of spectral parameters for each source and activity state is reported and discussed in greater detail in Appendix \ref{app:spectral-params}.

\subsection{Modelling of ALP-photon oscillations}
\label{sec:alp-photon}
The propagating ALP-photon beam can be described as a three-state quantum system, representing the two photon polarizations and the ALP state, obeying a Schr\"odinger-like equation \cite{raffelt1988}. However, in the general case of an unpolarized beam (as for gamma rays), the system is more correctly represented by a $3\times 3$ density matrix $\rho$ satisfying the von Neumann-like equation \cite{deangelis2011,galanti2022}
\begin{equation}
    \label{eq:photon-alp}
    i\dv{\rho}{y}=[\rho, \mathcal{M}_0(E)]\,,
\end{equation}
where $y$ is the propagation direction and $\mathcal{M}_0(E)$ is an energy-dependent matrix whose elements account for photon dispersion, absorption and for ALP-photon mixing in external magnetic fields. For a given magnetic field configuration, this equation can be solved iteratively over discrete magnetic domains by the method of transfer matrices \cite{mirizzi2008,bassan2010,deangelis2011,galanti2022}. Given the initial state $\rho_i$ at position $y_0$ and the total transfer matrix $\mathcal{U}(E;y,y_0)$, the probability of finding the system in the final state $\rho_f$ at position $y$ is expressed as 
\begin{equation}
    P_{if}(E;y) = \Tr[\rho_f\,\mathcal{U}(E;y,y_0)\,\rho_i\,\mathcal{U}^\dagger(E;y,y_0)]\,.
\end{equation}
In particular, for an initial unpolarized photon state $\rho_0=(1/2)\diag(1,1,0)$ and a final photon state of either polarization $\rho_1=\diag(1,0,0)$ or $\rho_2=\diag(0,1,0)$, the \textit{photon survival probablity} is given by
\begin{align}
\label{eq:pgg}
    \Pgg(E) = \Tr[(\rho_1+\rho_2)\,\mathcal{U}(E;y,y_0)\,\rho_0\,\mathcal{U}^\dagger(E;y,y_0)]\,.
\end{align}

In this work, the calculations outlined above are performed using the gammaALPs v0.3.0\footnote{\href{https://github.com/me-manu/gammaALPs}{https://github.com/me-manu/gammaALPs}} Python package \cite{meyer2021}. This package defines a {\tt ModuleList} class where different magnetic field environments can be added in sequence. The {\tt ModuleList} can then be run to automatically solve eq. \eqref{eq:photon-alp} over the line of sight from a given source and obtain $\Pgg(E)$ following eq. \eqref{eq:pgg}, for initial state $\rho_0$ and fixed ALP parameters $(m_a,\gag)$.

In order to compute the photon survival probabilities for our selected blazars, we consider  ALP-photon mixing in their jets' magnetic fields, modelled with helical and tangled components as in ref. \cite{davies2021}. The model is based on the blazar jet structure described by Potter and Cotter \cite{potter2013a,potter2013b,potter2013c,potter2013d} and implemented in the gammaALPs class {\tt MixJetHelicalTangled}. The parameters of each source are adapted from Figure 7 in ref. \cite{potter2015}\footnote{We note that ref. \cite{potter2015} employs the same source nomenclature used in the the \textit{Fermi}-LAT Bright Source List (0FGL) \cite{abdo2009,abdo2010}, and hence the source PKS~2155$-$304 appears listed as J2158.},
except for the free electron density normalization which is taken from Table 5 in ref. \cite{tavecchio2010}. Additionally, we include photon-ALP mixing in the Milky Way's galactic magnetic field as modelled in ref. \cite{jansson2012} and EBL absorption from ref. \cite{dominguez2011}. These propagation environments are implemented, respectively, in the classes {\tt MixGMF} and {\tt OptDepth} (from the ebltable package\footnote{\href{https://github.com/me-manu/ebltable}{https://github.com/me-manu/ebltable}}). 
Finally, a possible contribution from the intergalactic magnetic field is usually modeled as a randomly oriented field with present-day intensity $B_0\lesssim1\,{\rm nG}$ over a coherence length $l_0\lesssim1\,{\rm Mpc}$ \cite{ade2016}. While the effect of this field is not necessarily small (e.g \cite{deangelis2007,abramowski2013}), we decided to neglect it owing to the large uncertainty in its structure and magnitude, in line with previous works \cite{meyer2014b,ajello2016,abdalla2021,abe2024}.

\begin{figure*}
    \centering
    \includegraphics[width=0.49\linewidth]{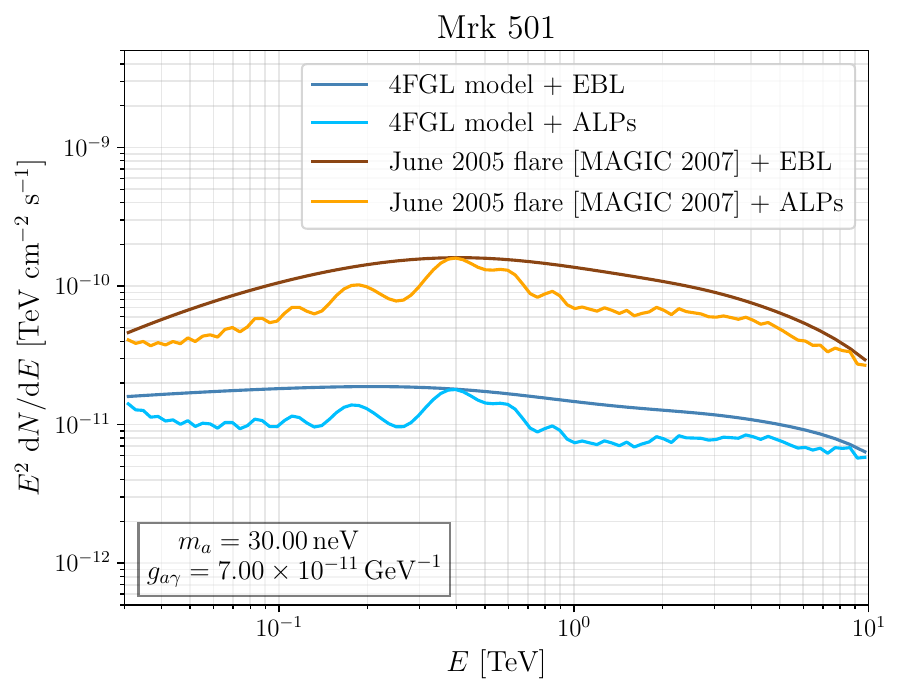}
    \includegraphics[width=0.49\linewidth]{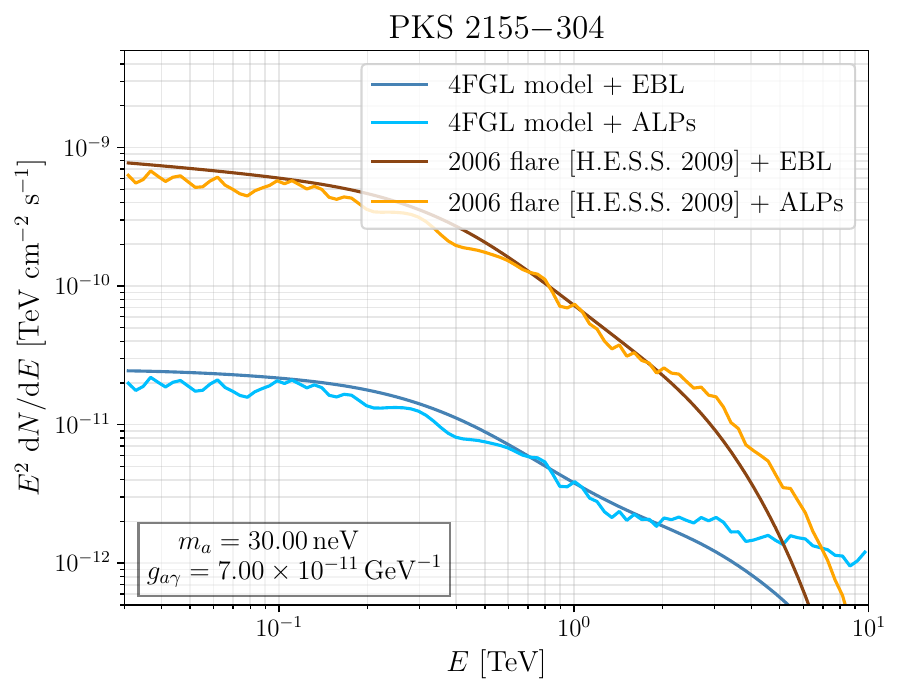}
    \caption{ALP effects on the baseline and flaring spectral shapes chosen for the sources in this work. For large enough values of $\gag$, both spectral distortions (``wiggles'') and a hardening at high energies are visible}
    \label{fig:alp-spectra}
\end{figure*}

In the presence of ALP-photon mixing, the intrinsic and observed fluxes are thus related by
\begin{equation}
\label{eq:obs-spectrum-alp}
    \phi_{\rm obs}(E)=\phi_{\rm int}(E)\Pgg(E)\,,
\end{equation}
where $\Pgg$ reduces to the EBL absorption factor $e^{-\tau}$ for $\gag=0$. Figure \ref{fig:alp-spectra} shows examples of the ALP effects on the baseline and flaring spectra of the blazars studied in this work.

\subsection{Data simulation and analysis}
\label{sec:simulations}
CTAO observations are simulated using the Gammapy v1.0.1 Python package \cite{gammapy2023,gammapy-1.0.1}. For each source, we generate 50 hours of mock data for the baseline state and 5 hours (corresponding to about a night of observations) for the flaring state. We assume observations to be taken in the ON-OFF mode, i.e. by pointing the telescopes at an offset from the source (the ON region) and estimating the background from a number of OFF regions symmetrically distributed around the center of the field of view. For this setup, we consider an offset from the source of 0.7 degrees, taking ON and OFF regions with a radius of 0.2 degrees. We choose 3 OFF regions, or equivalently, an acceptance ratio $\alpha=1/3$. The source coordinates are implemented in the {\tt SkyCoord} and {\tt SourceCatalog4FGL} classes, respectively from the Astropy \cite{astropy2013,astropy2018,astropy2022} and Gammapy\footnote{Note that Gammapy v1.3 \cite{gammapy-1.3} is necessary to load the 4FGL-DR4 catalog.} packages. The parameters of flaring spectra are taken directly from refs. \cite{albert2007,aharonian2009}, and the redshift parameters reported in TeVCat\footnote{\href{https://www.tevcat.org/}{https://www.tevcat.org/}} \cite{wakely2008} are used. 

\begin{table*}[t]
    \centering
    \caption{Summary table of the astrophysical parameters of the selected blazars. The gammaALPs parameters are adapted from refs. \cite{potter2015} and \cite{tavecchio2010}. Additionally, the parameters {\tt rvhe} and {\tt r\_T} are used, both set equal to {\tt r\_0} and taken to be $\sim100$ times the size of the transition region $R_T$. Unspecified parameters are set to their default values}
    \label{tab:astrophysical-params}
    \adjustbox{max width=\textwidth}{
    \begin{tabular}{lrrrrrrrrrr}
      \toprule
      \multirow{2}{*}{Source} & \multirow{2}{*}{$z$} & \multirow{2}{*}{RA [deg]} & \multirow{2}{*}{Dec [deg]} & \multicolumn{7}{c}{gammaALPs parameters} \\
      & & & & \texttt{r\_0} [pc] & \texttt{B0} [G] & \texttt{gmax} & \texttt{gmin} & \texttt{n0} [cm$^{-3}]$ & \texttt{rjet} [pc] & \texttt{alpha} \\
      \midrule
      Mrk 501      & 0.034 & 253.47 & 39.76 & 0.36 & 0.81 & 9  & 2  & $4.5\times 10^4$ & $3.2\times 10^3$ & 1.68 \\
      PKS 2155$-$304 & 0.116 & 329.72 & $-30.23$ & 0.33 & 0.82 & 15 & 7  & $1.65\times 10^4$ & $3.2\times 10^3$ & 1.70 \\
      \bottomrule
    \end{tabular}
    }
\end{table*}

\begin{table*}[t]
    \centering
    \caption{Observation times, IRF file names and reference spectra for the selected blazars}
    \label{tab:observation-params}
    \adjustbox{max width=\textwidth}{
    \begin{tabular}{lrll}
      \toprule
      Source & Livetime [h] & IRFs & Reference spectrum \\
      \midrule
      Mrk 501 (baseline)      & 50 & {\tt Prod5-North-20deg-AverageAz-4LSTs09MSTs.180000s} & LP \cite{4fgl-dr4} \\
      PKS 2155$-$304 (baseline)      & 50 & {\tt Prod5-South-20deg-AverageAz-14MSTs37SSTs.180000s} & LP \cite{4fgl-dr4} \\
      Mrk 501 (flaring)      & 5 & {\tt Prod5-North-20deg-AverageAz-4LSTs09MSTs.18000s} & LP \cite{albert2007} \\
      PKS 2155$-$304 (flaring)      & 5 & {\tt Prod5-South-20deg-AverageAz-14MSTs37SSTs.18000s} & ECPL \cite{aharonian2009} \\
      \bottomrule
    \end{tabular}
    }
\end{table*}

\begin{figure*}
    \centering
    \includegraphics[width=0.49\linewidth]{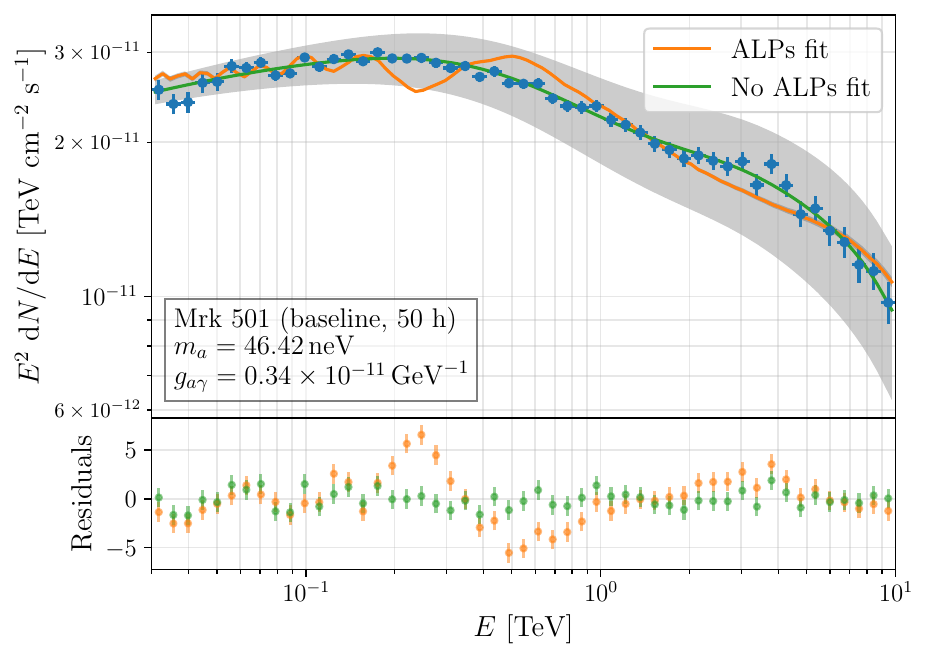}
    \includegraphics[width=0.49\linewidth]{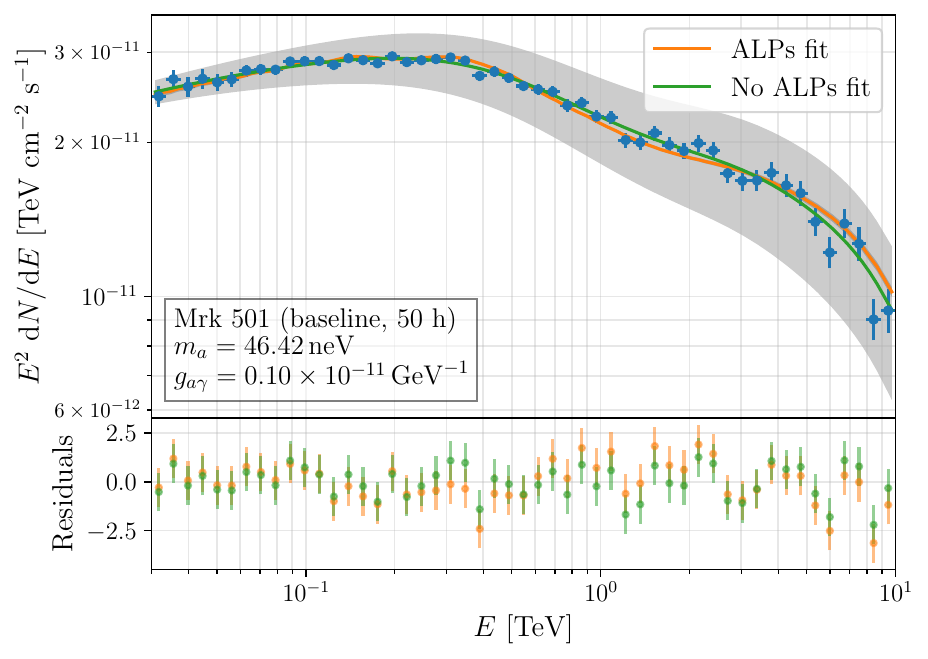}
    \caption{Example simulated SEDs for two 50-hour observations of the Mrk 501 baseline state. Spectral fits are shown for the ALP-less case and for two different ALP parameter sets. The residuals for both cases are computed as $(f-\phi)/\sigma_f$, where $f$ is the measured flux with uncertainty $\sigma_f$ and $\phi(E)$ is the model prediction (see ref. \cite{abdalla2021})}
    \label{fig:example-simulations}
\end{figure*}

We subdivide the measured energy axis with 20 bins per energy decade, in order to resolve the fine spectral irregularities that may be introduced by ALP effects. For each energy bin $i$, the expected number of signal counts $\mu_i$ is obtained by convolving the observed flux $\phi_{\rm obs}$ with the instrument response functions (IRFs), and then integrating over the energy bin width and observation time \cite{abdalla2021}. For our simulations we use the publicly available CTAO prod5 v0.1 IRFs \cite{ctao_irfs_prod5}, computed for the CTAO ``Alpha'' configuration described in Section \ref{sec:intro}. These include effective area, point spread function (PSF) and energy dispersion, as well as the background rates that can be used to compute the expected number of background counts per energy bin $b_i$. Different IRF files are available for 0.5, 5 and 50 hours of observations, both for the Northern and Southern Arrays in distinct subarray configurations, and for different values of zenith and azimuth angles. We simulate observations of Mrk~501 and PKS~2155$-$304 with the Northern and Southern Arrays, respectively, assuming all observations to be taken at a zenith angle of 20 degrees while averaging over the azimuth angle. 
The astrophysical and observational parameters used in simulating the datasets, as well as the chosen IRFs and spectral models, are summarised in Tables \ref{tab:astrophysical-params} and \ref{tab:observation-params}.

Datasets are simulated by sampling the signal and background counts in each energy bin from Poisson distributions, with average values $\mu_i +\alpha b_i$ in the ON region and $b_i$ in the OFF regions respectively, using the {\tt fake()} method provided by the {\tt Dataset} class in Gammapy. Example simulated SEDs for the baseline state of Mrk 501 are shown in Figure \ref{fig:example-simulations}.

\section{Derivation of ALP sensitivity with the likelihood-ratio test}
\label{sec:lr-limits}
Before presenting our method based on machine learning classifiers, we derive ALP limit projections using the standard LRT statistic, in order to introduce the general setup and to have a benchmark for our later results. Moreover, the results obtained with this approach can be directly compared with previous publications, where similar techniques were applied.

We consider an array of $10\times 10$ logarithmically spaced points over the ALP parameter space, with $m_a\in[0.1, 1000]\,{\rm neV}$ and $\gag\in[0.03,7]\times 10^{-11}\,{\rm GeV}^{-1}$. The choice of the mass interval is commonly adopted in similar studies, and roughly corresponds to the region where the critical energy for the strong mixing regime falls within the CTAO energy range \cite{meyer2014a}. Such energy value is defined as that above which the ALP-photon oscillation probability in a constant magnetic field becomes maximal and almost independent of energy \cite{meyer2014a,mirizzi2008,bassan2010,deangelis2011,galanti2022}. On the other hand, the upper bound on $\gag$ is motivated by the most recent experimental limit set by the CAST helioscope \cite{altenmuller2024}.

Given the intrinsic spectrum of each source, we compute the photon survival probability $\Pgg$ for each $(m_a,\gag)$ pair with gammaALPs as described in Section \ref{sec:alp-photon}, and obtain the expected observed spectrum as in eq. \eqref{eq:obs-spectrum-alp}. A likelihood function
\begin{equation}
    \mathcal{L}(m_a, \gag, \mu, b, \theta|D)=\prod_i\mathcal{L}(m_a, \gag, \mu_i, b_i, \theta|D_i)
\end{equation}
is defined, where each $i$-th energy bin contributes with a Poissonian term of the form
\begin{align}
    \begin{aligned}
        \mathcal{L}(m_a, &\gag, \mu_i, b_i, \theta|D) = \\ 
    &\frac{(\mu_i+\alpha\,b_i)^{n_{{\rm ON},i}}}{n_{{\rm ON},i}!}e^{-(\mu_i+\alpha\, b_i)}\frac{b_i^{n_{{\rm OFF},i}}}{n_{{\rm OFF},i}!}e^{-b_i}
    \end{aligned}
\end{align}
for a dataset $D$ with $n_{\rm ON}=\sum_i n_{{\rm ON},i}$ and $n_{\rm OFF}=\sum_i n_{{\rm OFF},i}$ photon counts in the ON region and in all OFF regions, respectively. In the former expression, $\mu_i$ is the expected signal photon count according to the ALP model with parameters $(m_a, \gag)$, while $b_i$ is the expected background photon count in the OFF regions. The expected signal count $\mu_i$ depends on the SED nuisance parameters (amplitude, spectral indices and cutoff energy), collectively denoted by $\theta$. The nuisance parameters $\theta$ and $b_i$, and consequently $\mu_i$, are set to the values $(\hat{\mu_i},\hat{b_i},\hat{\theta})$ that maximise the likelihood for a given choice of $(m_a, \gag)$ and dataset $D$. This procedure, called likelihood profiling, is equivalent to fitting the spectral model to the dataset.

Following ref. \cite{abe2024}, we then define the test statistic
\begin{equation}
\label{eq:ts-lr}
    {\rm TS}(m_a, \gag|D) =
    -2\ln\frac{\mathcal{L}(m_a, \gag, \hat{\mu},\hat{b},\hat{\theta}|D)}{\hat{\mathcal{L}}(D)}\,,
\end{equation}
where $\hat{\mathcal{L}}(D)$ is the maximum likelihood for the given dataset over the considered ALP parameter space. In practice, maximizing the likelihood is equivalent to minimizing the quantity
\begin{align}
    \begin{aligned}
        {\rm WSTAT} =&  2\sum_i [\mu_i+(\alpha+1)b_i\\
        &-n_{{\rm ON},i}\ln(\mu_i+\alpha b_i)-n_{{\rm OFF},i}\ln b_i]\,,
    \end{aligned}
\end{align}
which corresponds to $-2\ln\mathcal{L}$ after neglecting terms that do not depend on $\mu_i$ or $b_i$. This quantity is computed and returned by Gammapy when fitting each spectral model to the dataset $D$. Therefore, eq. \eqref{eq:ts-lr} is equivalent to
\begin{align}
    \begin{aligned}
        \label{eq:ts-wstat}
    {\rm TS}(m_a, &\gag|D) = \\& {\rm WSTAT}(m_a, \gag|D)-{\rm WSTAT}_{\rm min}(D)\,.
    \end{aligned}
\end{align}
Intuitively, this TS can be seen as a measure of the difference between any ALP model $(m_a,\gag)$ and the one that best fits the data.

\begin{figure}
    \centering
    \includegraphics[width=\linewidth]{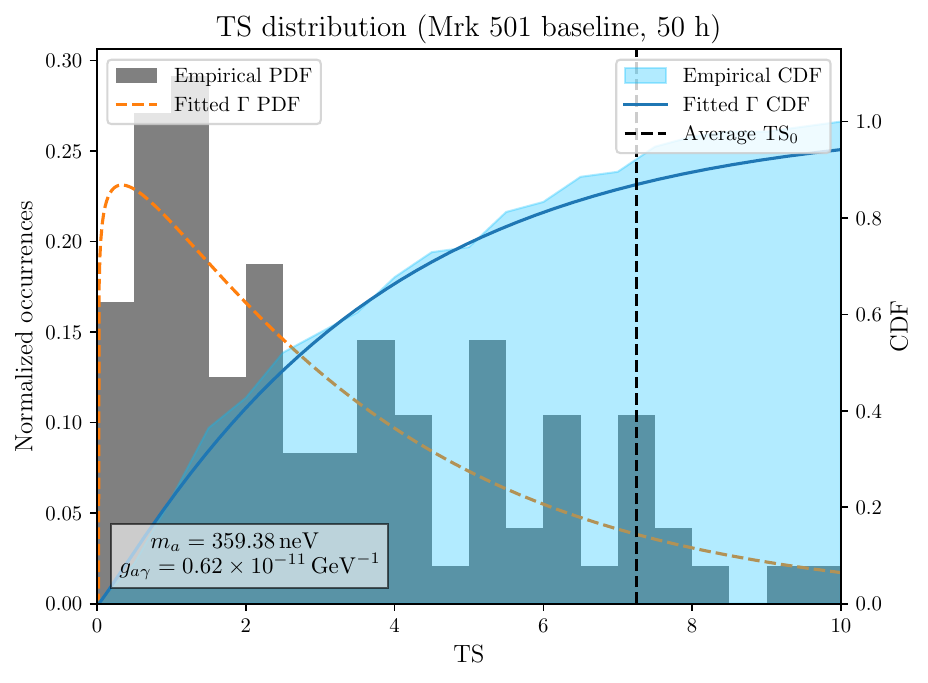}
    \caption{Example likelihood-ratio TS distribution obtained for the baseline state of Mrk~501}
    \label{fig:example-ts-distribution-std}
\end{figure}

For each source (considered in both baseline and flaring states) we simulate 100 observations without including ALP effects, following the prescriptions outlined in Section \ref{sec:simulations}, and compute the average TS for each $(m_a,\gag)$ pair. 
In order to obtain the statistical significance for excluding the ALP hypothesis with parameters $(m_a,\gag)$, one then needs to compare the obtained average, denoted with ${\rm TS_0}(m_a,\gag)$, against the TS distribution expected under the ALP hypothesis. As has been argued several times, Wilks' theorem \cite{wilks1938} cannot be applied to the case of ALPs \cite{meyer2014b,abe2024,abdalla2021}, and thus the shape of the TS distribution cannot be assumed a priori. Therefore, for each $(m_a, \gag)$ point and each considered source, we simulate 100 datasets including ALP effects and compute the corresponding TS values according to eq. \eqref{eq:ts-wstat}. An example of a TS distribution obtained in this way is shown in Figure \ref{fig:example-ts-distribution-std}. Following ref. \cite{abe2024}, we fit this distribution with a Gamma probability distribution function (PDF), given by
\begin{equation}
    f(x, \alpha, \beta) = \frac{\beta^\alpha x^{\alpha-1}e^{-\beta x}}{\Gamma(\alpha)}\,,
\end{equation}
where $\alpha,\beta>0$ and $\Gamma(\alpha)$ is the gamma function
\begin{equation}
    \Gamma(\alpha)=\int_0^\infty t^{\alpha-1}e^{-t}\dd{t}\,.
\end{equation}
The confidence level (CL) for excluding the ALP hypothesis $(m_a, \gag)$ is then given by the cumulative distribution function (CDF) evaluated at ${\rm TS_0}(m_a, \gag)$, i.e.
\begin{equation}
\label{eq:cl}
    {\rm CL}={\rm CDF}({\rm TS_0})\equiv\int_0^{{\rm TS_0}}f(x,\alpha,\beta)\dd{x}\,. 
\end{equation}
Finally, the exclusion significance is expressed as the equivalent number of standard deviations ($z$-score)
\begin{equation}
\label{eq:z-score}
    z = \sqrt{2}\,{\rm Erf^{-1}(CL)}\,,
\end{equation}
where ${\rm Erf}^{-1}(x)$ is the inverse of the error function
\begin{equation}
    {\rm Erf}(x)=\frac{2}{\sqrt{\pi}}\int_0^x e^{-t^2}\dd{t}\,.
\end{equation}
All the statistical computations discussed above were performed using the Scipy Python package \cite{virtanen2020}.

Intuitively, it can be understood that for any $(m_a,\gag)$ pair the TS distribution will generally be concentrated around small values: since the best-fitting ALP model will most likely be close to the model that was used to generate the datasets, the difference between the two log-likelihood values (measured by the TS) should be small and only due to statistical fluctuations. Conversely, the average of ALP-less datasets may present much larger ${\rm TS_0}$ values for certain $(m_a,\gag)$ pairs, which in turn would allow them to be excluded. We stress here that, since we are testing datasets \textit{without} ALP effects and we are computing the significance for excluding the ALP hypothesis, our results should be interpreted as an estimate of CTAO's capability of constraining ALPs (i.e., placing upper limits on their parameter space) rather than discovering them.

\section{Derivation of ALP sensitivity with machine learning classifiers}
\label{sec:ml-limits}
\subsection{Machine learning classifiers}
In this Section we give a brief review on the topic of ML classifiers, which we later apply in the derivation of upper limits on the ALP parameter space.

A basic example of a ML classifier is the \textit{binary decision tree} (BDT). A BDT can act on a dataset\footnote{In machine learning literature the term ``sample'' would be used, whereas ``dataset'' would denote a set of many samples.}, identified by a vector of features $\vb{X}=(X_1,\dots,X_n)$, and distinguish whether it belongs to one of two classes by operating sequential binary splits on the features' values. The splitting thresholds at each step (node) are referred to as the parameters of the tree; on the other hand, quantities such as the tree depth, maximum number of splits, etc. are called hyperparameters. During the ``supervised training'' of the BDT, the splits are chosen in order to minimize some loss function, by operating on datasets whose class is known a priori. Once the tree has been trained and its parameters have been fixed, it can be used to classify previously ``unseen'' test datasets. There are several methods to evaluate the performance of a ML classifier during the training phase, e.g. evaluating its accuracy and related metrics with train-test splitting and cross-validation. This ensures that a similar performance will be achieved on new data as well. A vast literature of textbooks and papers exists on the many scientific applications of ML classifiers and BDTs; see e.g. ref. \cite{raschka2017} for a basic introduction.

Random forests (RFs) are an example of \textit{ensemble} classifiers, since, as the name suggests, they are based on averaging a large number of binary decision trees. In order to avoid overfitting, i.e. classifiers performing ``too well'' on training data but badly on test data, it is necessary that the single BDTs stop after few splits (it is said that they are ``weak'' classifiers). BDTs, RFs and several other ML classification algorithms are implemented in the open-source Python package Scikit-learn \cite{pedregosa2012}. When one of these algorithms is trained to distinguish between two classes and then applied to some sample, it typically returns a set of scores indicating how likely the given sample is to belong to each of the classes.

Machine learning classifiers such as RFs find application in Cherenkov telescope data analysis, specifically in steps that involve the so-called gamma-hadron separation \cite{bock2004,albert2008,ohm2009}. Here, the atmospheric shower images are parametrized by a set of quantities (the Hillas parameters \cite{hillas1985}) which are then used as features to distinguish between showers initiated by hadrons and by gamma-ray photons.

XGBoost (eXtreme Gradient Boosting) \cite{chen2016} is a similar algorithm to the RF, which leverages a technique called \textit{gradient boosting} to efficiently minimise the loss function. Implementations of the XGBoost algorithm are available in several programming languages, including Python\footnote{\href{https://xgboost.readthedocs.io/en/stable/}{https://xgboost.readthedocs.io/en/stable/}}, and generally exhibit reasonably good performance ``out of the box'', with minimal tweaking compared to RFs. 
XGBoost is widely used in high energy physics due to its robustness and efficiency on structured, tabular data, especially in contexts where weak signals need to be disentangled from large backgrounds (e.g. ref. \cite{choudhury2024}). Applications to high-level astrophysical data are found e.g. in refs. \cite{chao2019,tolamatti2023}, where significant improvement in classification accuracy was found compared to other ML algorithms. This motivates our choice of employing XGBoost for ALP searches in the present work.

\subsection{Method}
\label{sec:ml-method}
In general, the effects of ALP-photon conversions on gamma-ray spectra are highly nonlinear and not easily parametrized. The energy and amplitude at which spectral features appear depend in a nontrivial way from the morphology and intensity of the magnetic fields encountered by the ALP-photon beam along the line of sight from Earth. It is therefore interesting to see how non-parametric statistical methods can address this task, such as those provided by ML classification algorithms. A first step in this direction is to verify that a ML-based method can reproduce the results obtained by the classical likelihood-ratio test (LRT) discussed above.

After being trained with a large amount of simulated spectra, both with and without ALP effects, a classifier is supposed to learn what the most important energy bins are to distinguish between the two classes, i.e., where the spectral oscillations and/or hardening are most relevant. This classification task crucially depends on the source under consideration, as well as on the ALP parameters $(m_a,\gag)$. Our approach was loosely inspired by that described in ref. \cite{day2020}. 

For each of our selected sources and activity states, we define a grid of classifiers over the points of the ALP parameter space described in Section \ref{sec:lr-limits}. 
The classifier corresponding to each $(m_a,\gag)$ pair is trained to distinguish between 20\,000 simulated datasets including ALP effects and 20\,000 only accounting for EBL absorption. The feature vector $\vb{X}$ used for training the classifiers is represented by the excess photon counts in each energy bin above background, which we assume to be already subtracted by conventional methods in the case of a real analysis.

These counts are then normalized to the $[0,1]$ interval, so that the only relevant characteristic for classifying the spectra is their \textit{shape}, given by the intrinsic differences between adjacent bins rather than the overall number of counts.
In order to allow for some variability in the intrinsic spectrum, avoiding to overfit the classifiers to a specific model, the training datasets are simulated by uniformly sampling values of each spectral parameter $p$ in the range $[p-\Delta p, p+\Delta p]$, according to the values reported in Table \ref{tab:spectral-params}. We have checked that the classifier performance only improves slightly by further increasing the number of training datasets.

The classifiers in the grid are based on the Python implementation of the XGBoost algorithm. 
The hyperparameters of the ML model were fixed to fiducial values, common to all sources and points in the ALP parameter space, motivated by the good observed performance of classifiers across the ALP grid. In order to assess this, we computed the values of several metrics (accuracy, precision, recall, F1 score) over the grid, showing that an accuracy higher than 90\% can be obtained over most of the relevant parameter space. These metrics were computed with Scikit-learn, using a 80-20 train-test split (for definitions, see e.g. ref. \cite{raschka2017}).

In all cases, an expected decrease in performance is observed at large $m_a$ and small $\gag$ values, in the lower right corner of the grid, where the spectral distortions due to ALPs are weaker and less distinguishable from the ALP-less case. As an example, the results obtained for the baseline state of Mrk~501 are shown in Figure \ref{fig:xgb-metrics}. The accuracy for all sources and states considered is reported in Appendix \ref{app:xgb-accuracy}, with the other metrics following a similar behaviour. In practice, the following hyperparameter values were used:
\begin{verbatim}
    n_estimators=100, max_depth=3,
    reg_lambda=500., reg_alpha=10.,
    eta=0.3
\end{verbatim}
A complete, point-by-point hyperparameter optimization will be considered in future developments of this work.

\begin{figure*}
    \centering
    \includegraphics[width=0.8\linewidth]{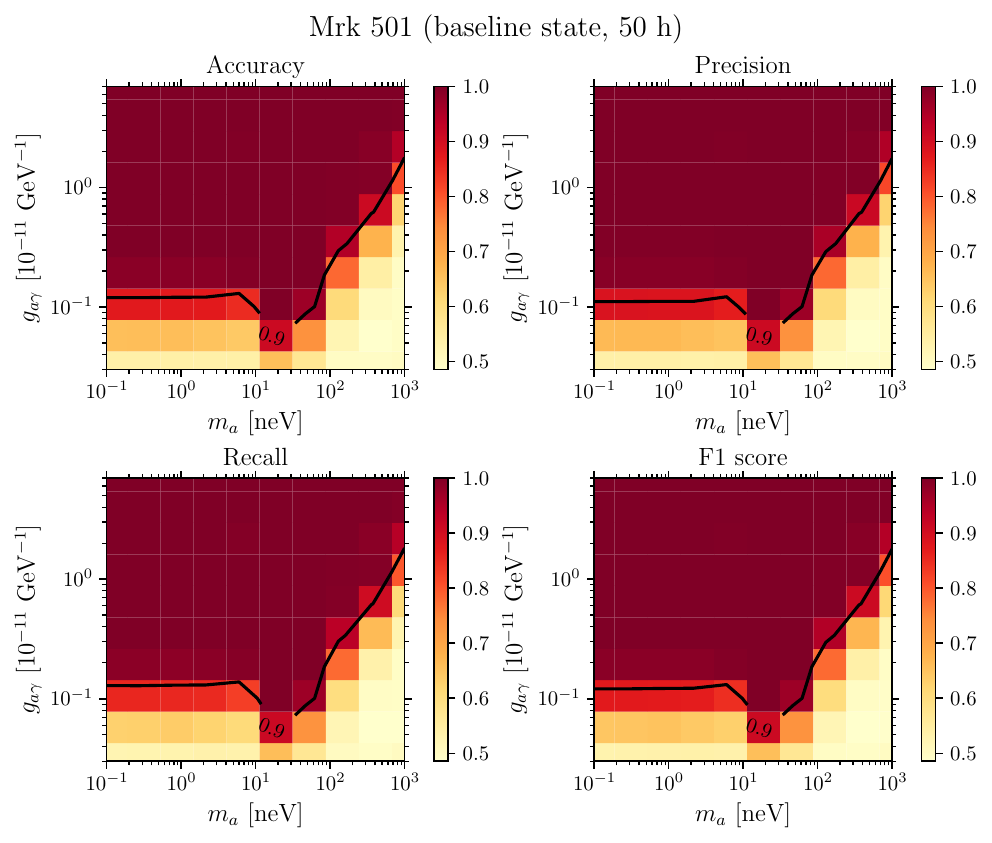}
    \caption{XGB metrics for the baseline state of Mrk~501 over the ALP parameter space}
    \label{fig:xgb-metrics}
\end{figure*}

Once the classifier grid for each source and spectral state has been trained, it is applied to 100 simulated datasets without ALP effects generated as described in Section \ref{sec:simulations}. For each dataset $D$, the classifier corresponding to each $(m_a,\gag)$ pair will return a score $p_{\rm ALP}(m_a,\gag|D)$ representing how likely the dataset is to be classified as ALP-like\footnote{We note that in general this score is not a probability in a strict statistical sense, unless it is appropriately calibrated.}. We then define the quantity
\begin{align}
\label{eq:ts-ml}
    \Pi(m_a,\gag|D)=1-p_{\rm ALP}(m_a,\gag|D)
\end{align}
and, similarly to the previous case, we denote the average value obtained from the 100 ALP-less datasets with $\Pi_0(m_a,\gag)$.

 We then search a $\Pi$ distribution for each $(m_a,\gag)$ pair, evaluating the $\Pi$ statistic in eq. \eqref{eq:ts-ml} for 2\,000 simulated ALP-like datasets. Since the value of $\Pi$ is always between 0 and 1, we fit the resulting distribution to a Beta PDF
\begin{equation}
    \label{eq:beta}
    f(x,\alpha,\beta)=\frac{\Gamma(\alpha+\beta)x^{\alpha-1}(1-x)^{\beta-1}}{\Gamma(\alpha)\Gamma(\beta)}\,,
\end{equation}
with $0\leq x\leq1$ and $\alpha,\beta>0$. Using this distribution, the confidence level for excluding the ALP hypothesis and the $z$-score are computed using eqs. \eqref{eq:cl} and \eqref{eq:z-score}, respectively. The quality of the fit is generally good for medium and low values of $\gag$, as can be seen from Figure \ref{fig:example-ts-distribution-ml}, but it worsens for higher ones. We have checked in Appendix \ref{app:fit-quality} that this does not significantly impact the main conclusions of this work. 

\begin{figure*}
    \centering
    \subfloat[\label{fig:example-ts-distribution-ml-good}]{
        \includegraphics[width=.48\textwidth]{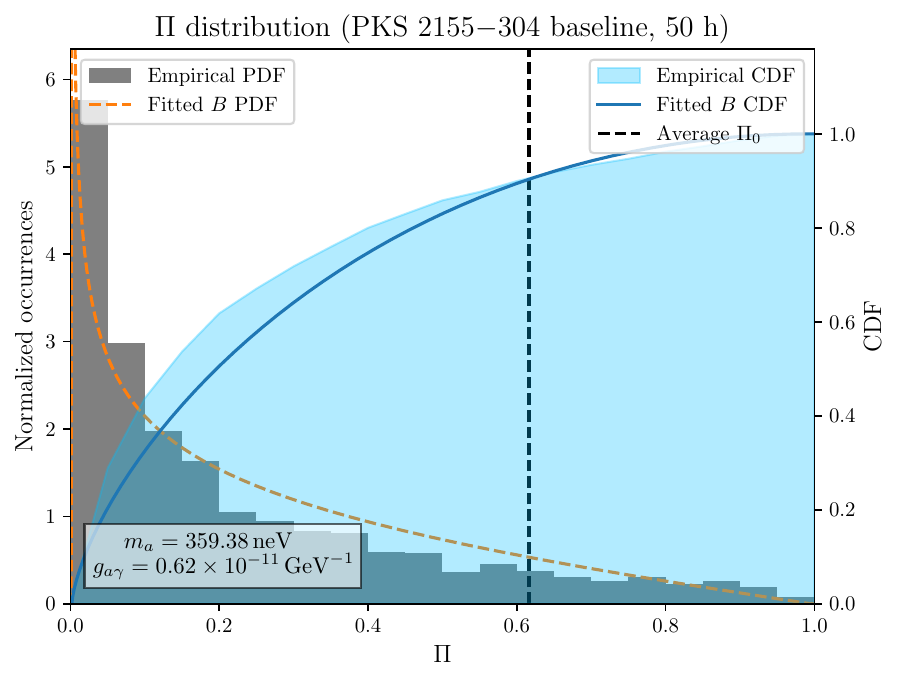}
        }
    \hfill
    \subfloat[\label{fig:example-ts-distribution-ml-bad}]{
        \includegraphics[width=.48\textwidth]{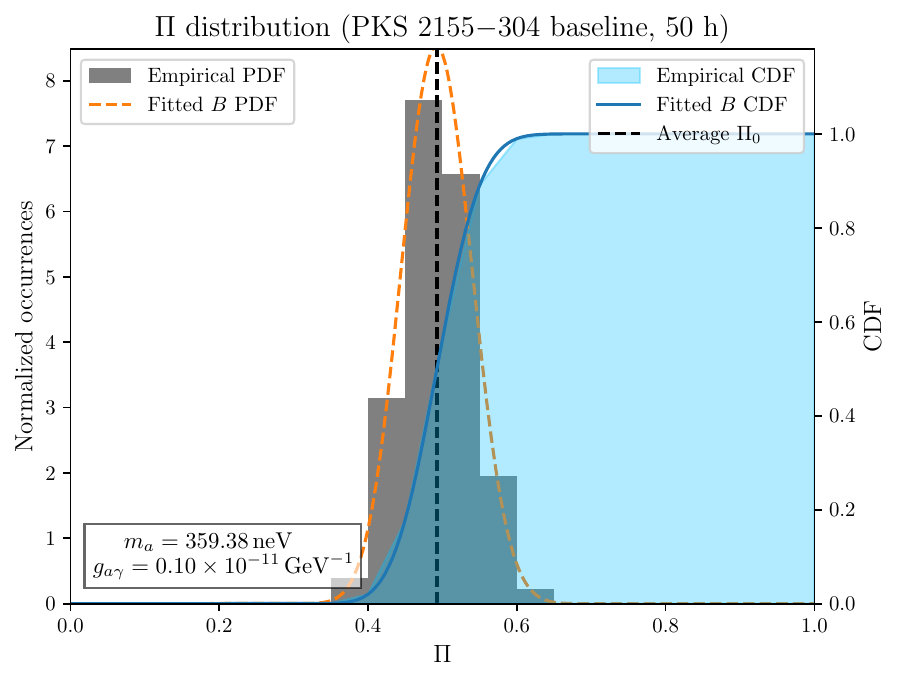}
        }
    \caption{Example distributions of the $\Pi$ statistic defined in eq. \eqref{eq:ts-ml}, comparing good and suboptimal performance of a classifier algorithm in different regions of the ALP parameter space}
    \label{fig:example-ts-distribution-ml}
\end{figure*}

\begin{figure*}
    \centering
    \includegraphics[width=0.49\linewidth]{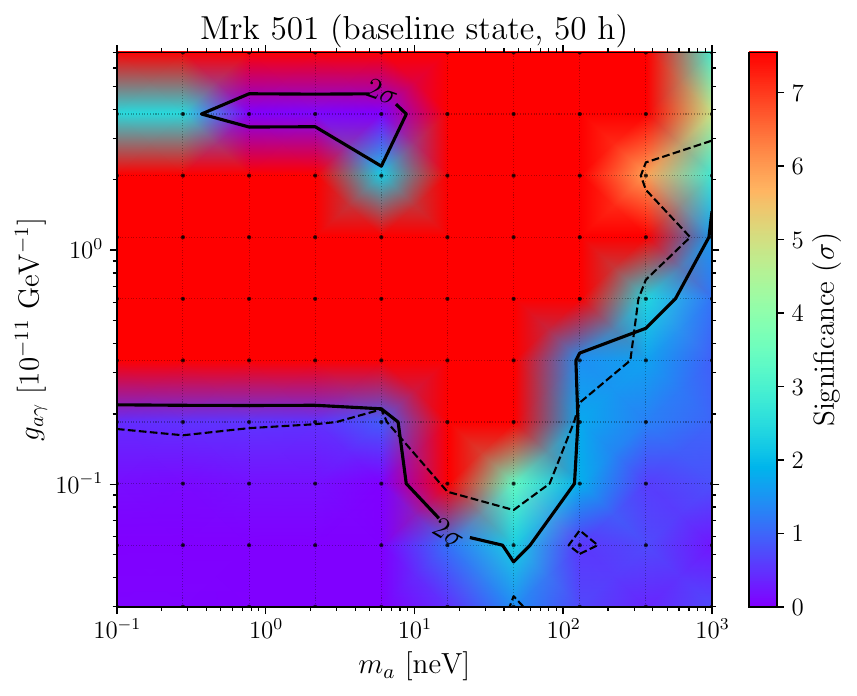}
    \hfill
    \includegraphics[width=0.49\linewidth]{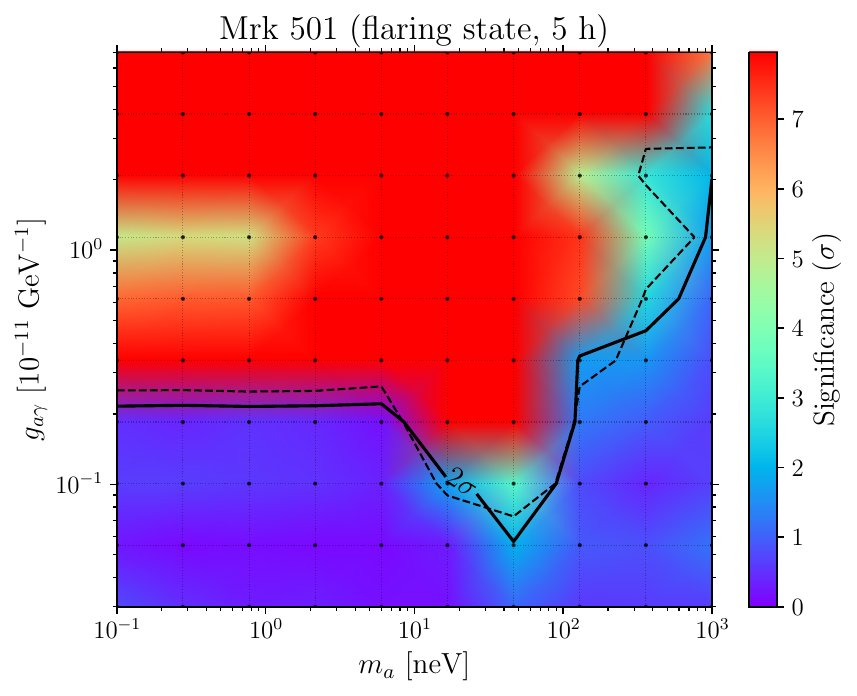}
    \hfill
    \includegraphics[width=0.49\linewidth]{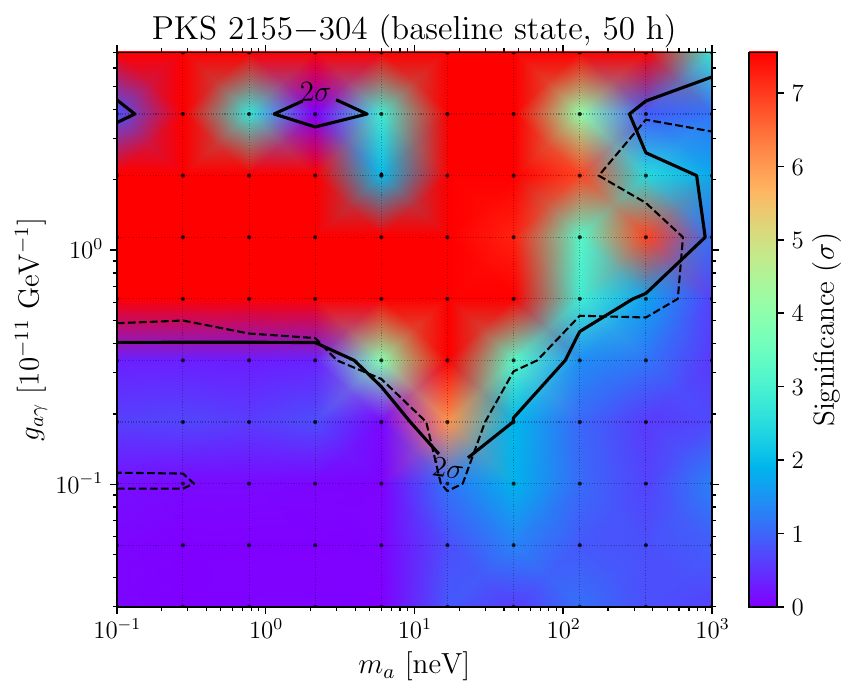}
    \hfill
    \includegraphics[width=0.49\linewidth]{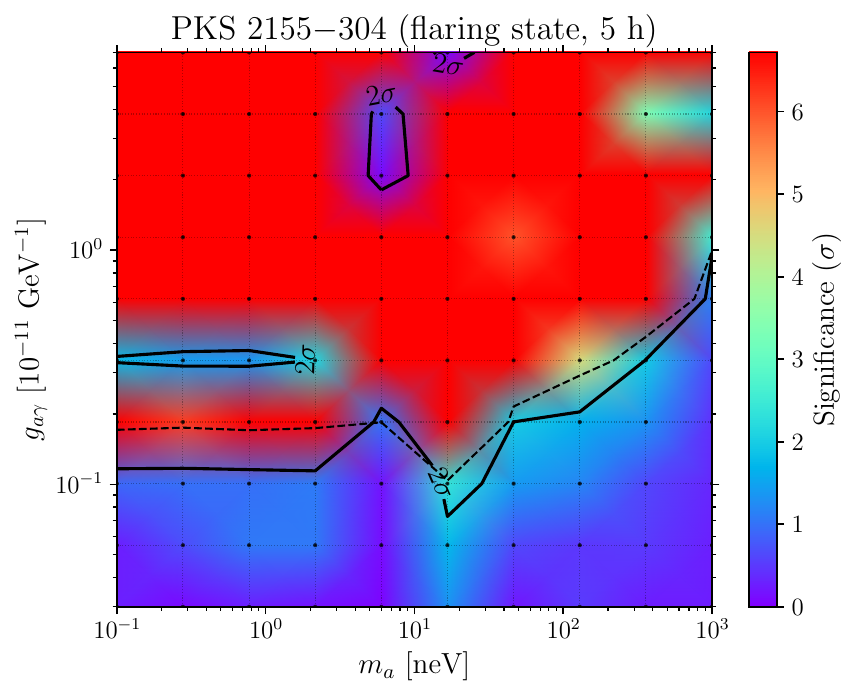}
    \hfill
    \caption{ALP exclusion significance computed with the ML-based method, with the $2\sigma$ contours highlighted (solid lines). The contours obtained at the same confidence level with the LRT method are shown as dashed lines for comparison. A linear interpolation was applied between the points in the parameter space for which the exclusion confidence level was computed, which are indicated as black dots}
    \label{fig:results-ml}
\end{figure*}

Clearly, for a well-performing classifier, the $\Pi$ distribution obtained in this way should be largely concentrated around small values, i.e. an ALP-like dataset is likely to be classified correctly. On the other hand, the $\Pi_0$ value from the ALP-less datasets should be quite high (close to 1), thus leading to exclusion of the ALP hypothesis. There are, however, some regions of the ALP parameter space where the spectra with and without ALP effects are hardly distinguishable, typically for large $m_a$ and small $\gag$ values. The accuracy of the classifiers in these cases is then very low (cf. Figure \ref{fig:xgb-metrics}), as can be seen from a $\Pi$ distribution peaked around 0.5, and no exclusion is expected. Examples of distributions corresponding to these two situations are shown in Figures \ref{fig:example-ts-distribution-ml-good} and \ref{fig:example-ts-distribution-ml-bad}, respectively.

\section{Discussion of results}
\label{sec:discussion}
Our main results are presented in Figure \ref{fig:results-ml}, where a color map is used to represent the exclusion significance obtained with the ML classifier method in units of Gaussian standard deviations over the ALP parameter space. We note that, for both methods, the exclusion significance may be infinite for those points where the confidence level computed by eq. \eqref{eq:cl} is equal to 1. For plotting, these values are clipped to the highest finite significance; see also Appendix \ref{app:fit-quality}.

The $2\sigma$ sensitivity obtained with the method of ML classifiers, shown as solid lines in Figure \ref{fig:results-ml}, is comparable with that resulting from the LRT (dashed lines). For all sources, the overall shapes and extensions of the projected exclusion regions are in agreement between the two methods, with occasional low-significance ``holes'' or high-significance ``islands'' in both sets of contours. Such features are common to other analyses of this kind \cite{ajello2016,abdalla2021} and do not significantly affect our main results.

The excluded region of the parameter space in the case of Mrk~501 is almost the same for both the baseline and flaring states. An explanation for this is the fact that the spectral shapes considered for the two states are fairly similar (see Fig. \ref{fig:alp-spectra}), and that the increased brightness of the flaring state is compensated by the longer exposure used for the baseline state. Given this, we can assume that the photon counts in the considered energy ranges are above the CTAO sensitivity for both states and are equally affected by ALPs, thus leading to similar exclusions in the parameter space. Conversely, PKS~2155$-$304 exhibits a spectral curvature and a stronger EBL absorption effect that significantly reduce the flux at higher energies. In this case, a flaring state can indeed bring the flux of higher-energy photons above the CTAO sensitivity, thus improving the constraining power of the observations. These qualitative considerations are confirmed by computing the Li-Ma significance $S$ of the signal \cite{li1983}. While for Mrk 501 both observations yield $S\sim300$, in the case of PKS~2155$-$304 a 5-hour flaring observation results in a significance $S\sim800$, compared to $S\sim200$ for a 50-hour observation in the baseline state.

These projections, representing the mean expected CTAO sensitivity to ALPs from each considered source, are obtained by averaging the results from many different simulated observations. A different approach for estimating such a mean sensitivity, employed in various studies on ALPs \cite{meyer2014a,meyer2014b,abdalla2021} and WIMPs \cite{acharyya2021,acharyya2023,abe2024b}, would be to use so-called ``Asimov datasets'' \cite{cowan2011} instead. These datasets are generated by simply setting the numbers of counts in the ON and OFF regions equal to their expected values, in order to eliminate Poissonian fluctuations and directly provide a median estimate of the sensitivity. We have checked explicitly that the contours obtained from a single Asimov dataset using the LRT method are very close to those resulting from averaging over many ``noisy'' datasets. On the other hand, we found that applying the ML classifiers to Asimov datasets causes a significant overestimation of the exclusion regions, as the total absence of Poissonian noise in photon counts almost always produces low $p_{\rm ALP}$ values even for extremely small $\gag$. In a realistic analysis, these points of the parameter space would not be excluded, as the inevitable photon noise could easily mimic ALP effects at very low couplings. We thus conclude that Asimov datasets cannot be reliably used in estimating ALP sensitivities with ML classifiers.

As a final remark, we note that these results may be affected by several systematic effects, such as the choice of energy binning, magnetic field parameters and EBL model. We discuss these effects in Appendix \ref{app:systematics}. We note, in particular, that recent measurements of the cosmic infrared background by the CIBER experiment might extend the allowed parameter space for ALPs to a wider region than what is expected from the commonly adopted EBL models \cite{kohri2017}.

\section{Conclusions}
\label{sec:conclusions}
In this work, we have estimated the capability of CTAO to place upper limits on the ALP parameter space by simulating observations of the blazars Mrk~501 and PKS 2155$-$304 with the Northern and Southern Arrays of the Observatory. The source selection and observation criteria were motivated by the AGN Key Science Project described in ref. \cite{ctascience2018}. Mock datasets for each source were prepared considering both long-term observations of baseline states, with spectra taken from the \textit{Fermi}-LAT 4FGL-DR4 source catalog \cite{4fgl-dr4}, and short flares based on previous TeV observations \cite{albert2007,aharonian2009}.

\begin{figure}
    \centering
    \includegraphics[width=\linewidth]{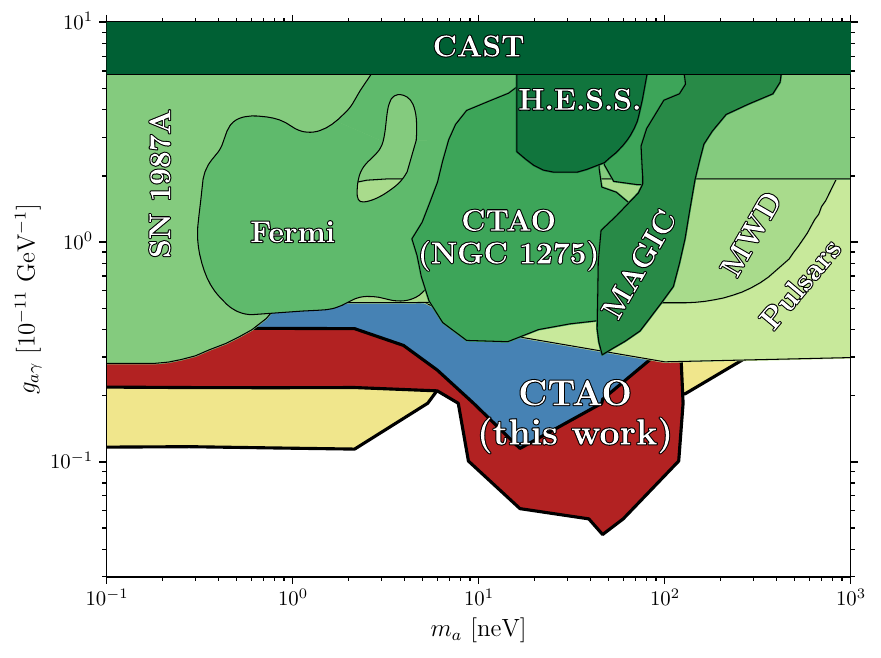}
    \caption{Comparison of the CTAO $2\sigma$ exclusion regions on the ALP parameter space obtained in this work with reference constraints from the literature (green). The projected sensitivities obtained from 50-hour observations of Mrk~501 (red) and PKS~2155$-$304 (blue) in their baseline states, as well as that from a 5-hour observation of a PKS~2155$-$304 flare (yellow) are shown. We do not explicitly include the results obtained from the Mrk~501 flare as they largely overlap with those from the baseline observation}
    \label{fig:limits-comparison}
\end{figure}

In Figure \ref{fig:limits-comparison} we compare the projections derived in this work with previously published constraints on the ALP parameter space\footnote{The reference limits are mostly taken from the online repository at ref. \cite{AxionLimits}.}. 
Most of these are computed at the $2\sigma$ confidence level \cite{abramowski2013,ajello2016,abdalla2021,altenmuller2024,dessert2022,noordhuis2023,manzari2024}, except for the limits by MAGIC which are reported at the $3\sigma$ level \cite{abe2024}. In particular, ref. \cite{abdalla2021} represents the only previous estimate of ALP sensitivity officially published by the CTAO Consortium, derived from simulated observations of the AGN NGC~1275 inside the Perseus galaxy cluster. 
With these targets, we expect that CTAO will be able to place limits on a much wider portion of previously unconstrained ALP parameter space in the $0.1\mbox{--}100\,\rm{neV}$ mass range. We finally remark that, while already competitive, the limits derived from a Southern source like PKS~2155$-$304 could be further improved with the foreseen addition of 2 LSTs to the Southern Array of CTAO, whose contribution would increase the sensitivity to low-energy spectral oscillations.

Astrophysical ALP searches are affected by large systematic errors. These mainly stem from the uncertainty in the modeling of the morphology and intensity of the magnetic fields along the line of sight from the source. The magnetic field structure in blazar jets, in particular, is almost unconstrained experimentally and can only be estimated from theoretical arguments, thus making ALP searches from blazars somewhat model-dependent. In the absence of a more precise determination of this structure, turning to different computational approaches that could detect ALP-induced spectral features in a model-independent fashion might be a viable strategy. To probe this possibility, we computed CTAO sensitivity projections by applying the classical likelihood-ratio test and a newly proposed method making use of machine-learning classifiers. While in the present context the classifier-based method uses the same modeling assumptions as the traditional method, thus not really solving the issue of systematic uncertainties, a promising result is shown by the fact that the sensitivities obtained by the two results broadly agree with each other, as shown in Figure \ref{fig:results-ml}. This is a first step in the direction of exploring different non-parametric methods for ALP searches in astrophysical data. Future developments of this concept will involve dedicated studies focused on the application of more complex and optimized algorithms, such as neural networks, with the aim of maximizing the attainable experimental sensitivity to ALPs while reducing the amount of model-dependent assumptions.

We conclude that observations of blazars with CTAO will be a powerful means to probe a significant portion of the parameter ALP space beyond the regions previously excluded by experiment, helping in guiding searches for these elusive particles as part of the AGN Key Science Project. To this aim, in this work we have introduced an alternative approach to constrain the ALP parameter space from binned gamma-ray data, based on the use of machine-learning classifiers, which offers promising results in good agreement with the standard likelihood-ratio test.

\paragraph{Acknowledgments}
We thank the CTAO internal reviewers, G. D'Amico, G. Galanti and M. Garczarczyk, for their insightful suggestions. F. S. would additionally like to thank A. Di Florio, F. Simone, F. Cuna, A. Mirizzi and K. Kohri for the comments they provided during various phases of this work. 

This work was conducted in the context of the CTAO DMEP working group. CTAO gratefully acknowledges financial support from the agencies and organizations listed at \href{https://www.ctao.org/for-scientists/library/acknowledgments/}{https://www.ctao.org/for-scientists/library/\\acknowledgments/}. F. S. acknowledges financial support by the EU funding program ``NextGenerationEU'' in the context of the PNRR-IR ``CTA+ - Cherenkov Telescope Array Plus'' program.

We would like to thank the computing centers that provided resources for the IRF generation, listed at \\\href{https://zenodo.org/records/5499840}{https://zenodo.org/records/5499840}.

\appendix

\section{Systematic effects}
\label{app:systematics}
In this Appendix we discuss how our results are affected by changes in energy binning, magnetic field strength or EBL model. While the former effects are similar between the sources considered in this work, we focus our discussion on PKS~2155$-$304 as it is the most strongly affected by EBL absorption.

\begin{figure*}
    \centering
    \subfloat[\label{fig:binning}]{
    \includegraphics[width=0.48\linewidth]{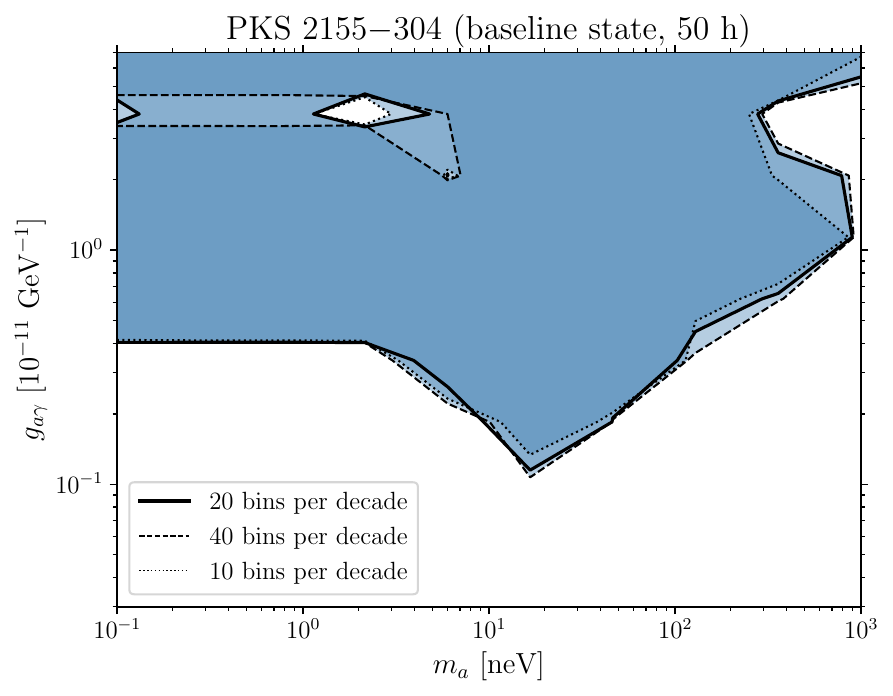}}
    \hfill
    \subfloat[\label{fig:ebl}]{
    \includegraphics[width=0.48\linewidth]{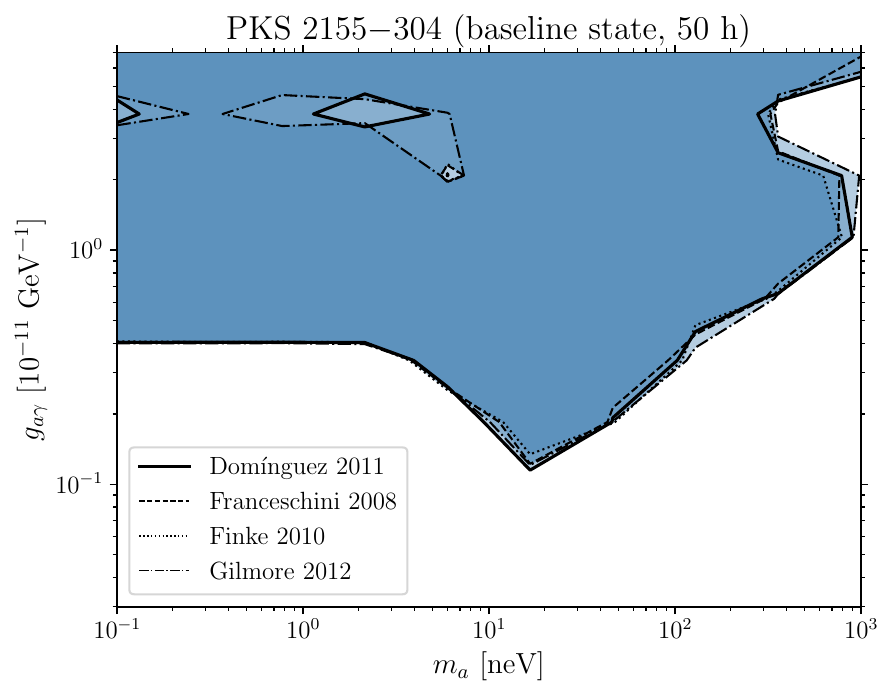}}
    \hfill
    \subfloat[\label{fig:mf-strength}]{
    \includegraphics[width=0.48\linewidth]{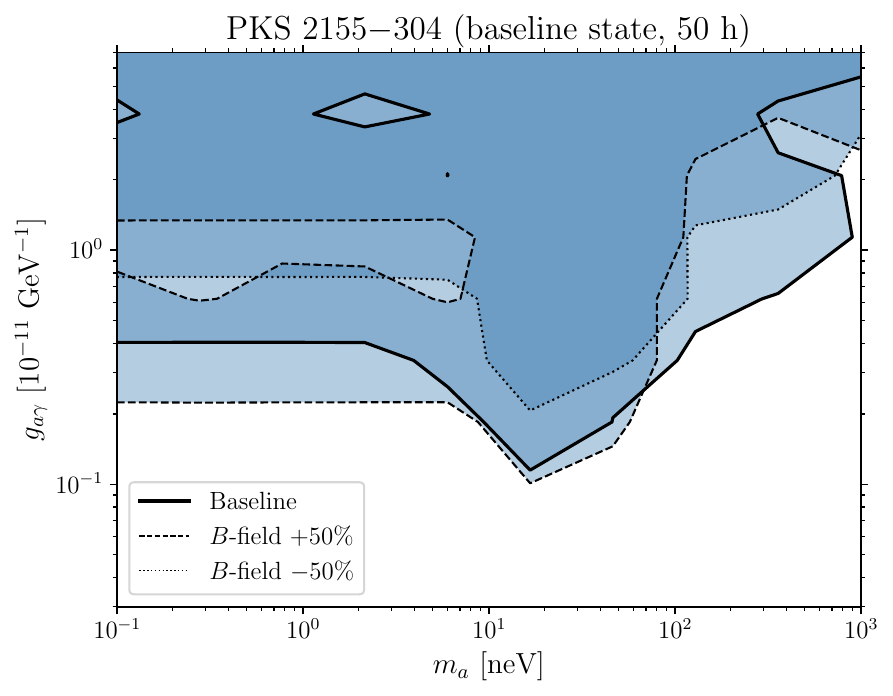}}
    \hfill
    \subfloat[\label{fig:training-strategy}]{
    \includegraphics[width=0.48\linewidth]{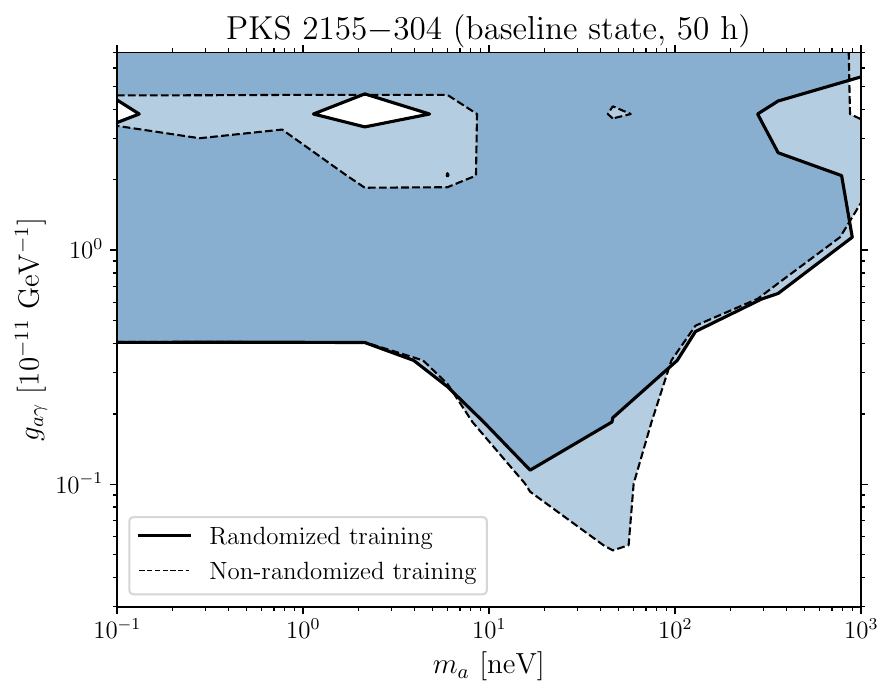}
    }
    \caption{CTAO sensitivity to the ALP parameter space obtained from a 50-hour simulated observation of PKS 2155$-$304 in its baseline state for different choices of the energy binning, EBL model, magnetic field strength and training strategy}
    \label{fig:systematics}
\end{figure*}

Figure \ref{fig:binning} shows the average $2\sigma$ sensitivity on the ALP parameter space from a 50-hour observation of PKS~2155$-$304 in its baseline state for different choices of the energy binning. We compare the results presented in the main text, obtained for 20 bins per decade, with those resulting from considering either half or double the number of bins per decade (10 or 40, respectively). The latter choice of binning, in particular, was employed in ref. \cite{abdalla2021}. While there is very little difference between these results, halving the number of bins seems to reduce the size of the hole-like features around $\gag\sim4\times10^{-11}\,{\rm GeV}^{-1}$, while doubling it turns the feature into a continuous band. A similar outcome is found by comparing the results from different EBL models \cite{franceschini2008,finke2010,gilmore2012} to the one used in the main text \cite{dominguez2011}, with the sensitivity to ALP parameters almost unaffected by this choice (Figure \ref{fig:ebl}).

Another source of systematic uncertainty is represented by the morphology and intensity of the jet magnetic field, which is poorly constrained experimentally. In this work we have used the fiducial model developed in ref. \cite{davies2021}, with the parameters of the Potter and Cotter jet model reported in ref. \cite{potter2015}. The same magnetic field geometry has already been used to derive constraints on the ALP parameter space \cite{davies2023,zhou2025}, and the relative impact of the model parameters on the mixing has been evaluated \cite{davies2021,gao2025}. To bracket our uncertainty on the magnetic field intensity in the emission region,
we consider in Figure \ref{fig:mf-strength} how our results are affected by changing the average magnetic field strength $B_0$ in the jet, either increasing it by 50\% ($B_0={1.23\,\rm G}$) or decreasing it by the same amount ($B_0={0.41\,\rm G}$) with respect to the value reported in Table \ref{tab:astrophysical-params} ($B_0={0.82\,\rm G}$). As the ALP effects on the photon spectrum depend on the product between the coupling constant $\gag$ and the magnetic field, it is easy to see that increasing the latter allows to set more stringent limits on the former, and vice versa. 
We can expect the projected limits to change by similar margins when varying other parameters, such as the distance of the emission region from the central black hole $r_0$. We remark that the Potter-Cotter model was successfully used to fit quasi-simultaneous broadband SEDs of several blazars during quiescent states \cite{potter2015}. However, these values are unlikely to remain constant while the source goes through different activity states. In future applications to real data, source variability should be accounted for by subdividing the data in different activity periods (e.g. using the method of Bayesian blocks \cite{abe2025}), and the parameters of the emission region should be estimated separately for each data set. In addition, parameters such as the size of the emission region should be consistent with the observed variability of the source. This approach would guarantee an accurate modeling of the source in different states, consequently reducing the uncertainty on the derived ALP limits.

Finally, in Figure \ref{fig:training-strategy} we compare our main results with those we find by not randomizing the intrinsic spectral parameters within their error bands in the training datasets (see Section \ref{sec:ml-method}) and only using their average values. As expected, the sensitivity derived in this way is stronger, as the ML classifiers become overfit to a specific spectral model, but less generalizable. On the other hand, in a ``randomized'' training, the variations on the spectral parameters should not be larger than a few percent in order to maintain a solid performance of the classifiers.

\section{Details on spectral models}
\label{app:spectral-params}
The parameters of each spectral model considered in this work are reported in Table \ref{tab:spectral-params}, where LP and ECPL refer to the log-parabola and power law with exponential cutoff models defined in eqs. \eqref{eq:lp} and \eqref{eq:ecpl}.
These models were implemented in Gammapy using the {\tt LogParabolaSpectralModel} and {\tt ExpCutoffPowerLawSpectralModel} classes, respectively. As discussed in the main text, the baseline spectra were taken from the \textit{Fermi}-LAT 4FGL-DR4 catalog \cite{4fgl-dr4}, while the flaring spectra were selected from previously published IACT observations. In particular, the EBL-corrected parameters of the Mrk~501 flaring spectrum are taken from Table 7 and Figure 19 of ref. \cite{albert2007}, referring to the flare of 30 June 2005. The PKS~2155$-$304 parameters are quoted from Table 3 in ref. \cite{aharonian2009} and correspond to the data taken with the lowest energy threshold ($\rm 200\,GeV$, labelled as {\tt T200}), corrected for EBL absorption.

\begin{table*}[]
    \centering
    \caption{Parameters of the spectral models used for the sources considered in this work. We note that the $\beta$ parameter for the Mrk 501 flaring state is scaled by a factor $\ln(10)$ with respect to ref. \cite{albert2007} after the conversion from base-10 logarithm to natural logarithm}
    \label{tab:spectral-params}
    \adjustbox{max width=\textwidth}{
    \begin{tabular}{*{8}{c}}
        \hline
         Source & Model & $\phi_0$ & $E_0$ & $\alpha$ & $\beta$ & $E_{\rm cut}$ & Ref.\\
         & Baseline & [$10^{-12}\,\rm{ MeV^{-1}\,cm^{-2}\,s^{-1}}$] & [MeV] & & & \\
         \hline
         Mrk 501 & LP & $3.76\pm0.05$ & 1535.09 & $1.75\pm0.01$ & $0.019\pm0.004$  & $-$ & \cite{4fgl-dr4}\\
         PKS 2155$-$304 & LP & $12.59\pm0.13$ & 1160.97 & $1.77\pm0.01$ & $0.042\pm0.004$ & $-$ & \cite{4fgl-dr4}\\
        \hline
        \hline
         & Flaring & [$10^{-10}\,\rm{TeV^{-1}\,cm^{-2}\,s^{-1}}$] & [TeV] & & &  [TeV] & \\
         \hline
         Mrk 501 & LP & $18.6\pm0.06$ & 0.3 & $1.73\pm0.06$ & $0.13\pm0.03$ & $-$ & \cite{albert2007} \\
         PKS 2155$-$304 & ECPL & $4.51\pm0.36$ & 1.0 & $2.16\pm0.05$ & $-$ & $1.74\pm0.24$ & \cite{aharonian2009}   \\
         \hline
    \end{tabular}
    }
\end{table*}

\begin{figure*}
    \centering
    \includegraphics[width=.8\textwidth]{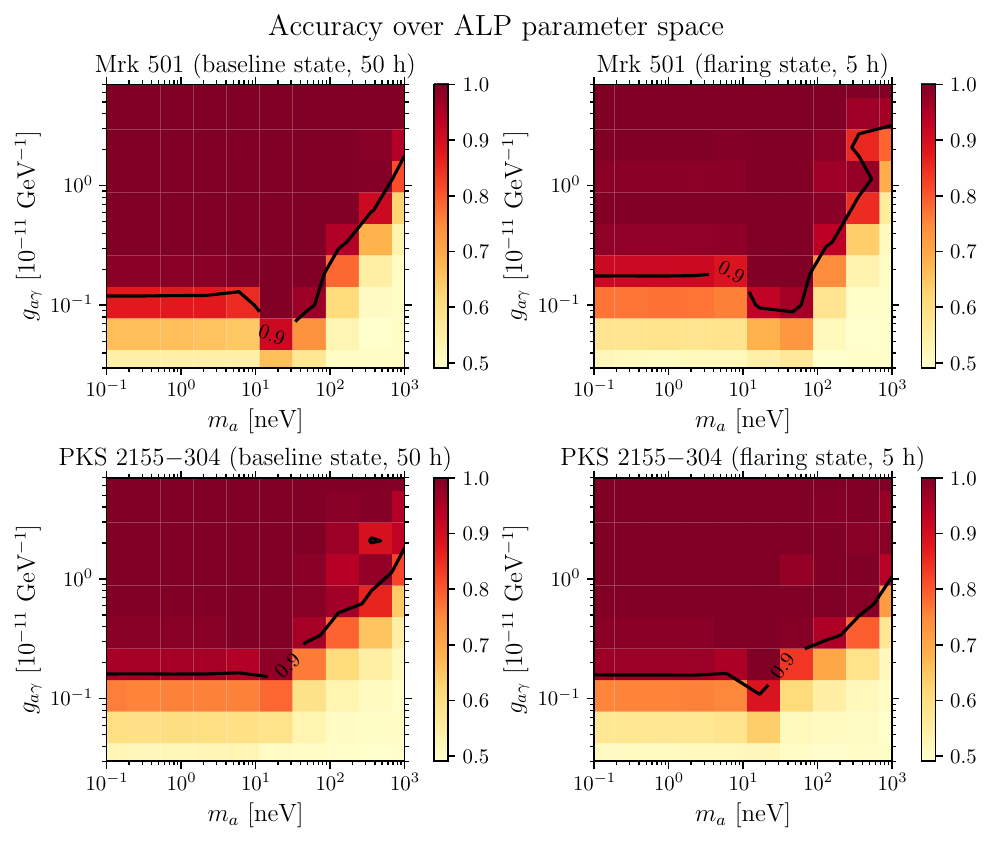}
    \hfill
    \caption{Accuracy of the XGBoost classifier grids defined over the ALP parameter space with the hyperparameters defined in Section \ref{sec:ml-limits}, with the $90\%$ contours highlighted}
    \label{fig:xgb-accuracy}
\end{figure*}

\section{Classifier accuracy over the ALP parameter space}
\label{app:xgb-accuracy}
In Figure \ref{fig:xgb-accuracy} we report the accuracy of the XGBoost classifiers defined over the ALP parameter space for each source, using the hyperparameters defined in Section \ref{sec:ml-limits}. As noted in the main text, we observe that the classifiers' performance is reasonably good (accuracy $\gtrsim90\%$) in the same regions of the parameter space where we expect CTAO to cast the strongest limits. All the other metrics that we have checked (precision, recall, F1 score) have almost the same contours as accuracy, as shown in Figure \ref{fig:xgb-metrics} for the baseline state of Mrk 501.

\section{Goodness of fit for the Beta distribution}
\label{app:fit-quality}
The example distributions of the $\Pi$ statistic shown in Figure \ref{fig:example-ts-distribution-ml} refer to cases where the accuracy of the classifiers is either worst or average. In these situations, the $\Pi$ values are more or less spread across the [0,1] interval and the resulting distribution is typically well fit by the Beta distribution of eq. \eqref{eq:beta}. This can be checked e.g. by applying the Kolmogorov-Smirnov (KS) test\footnote{\href{https://docs.scipy.org/doc/scipy/reference/generated/scipy.stats.kstest.html}{https://docs.scipy.org/doc/scipy/reference/generated/\\scipy.stats.kstest.html}}, which returns a large $p$-value when the data are compatible with the proposed distribution (i.e., the hypothesis that the data are described by said distribution is not rejected). 

\begin{figure}
    \centering
    \includegraphics[width=\linewidth]{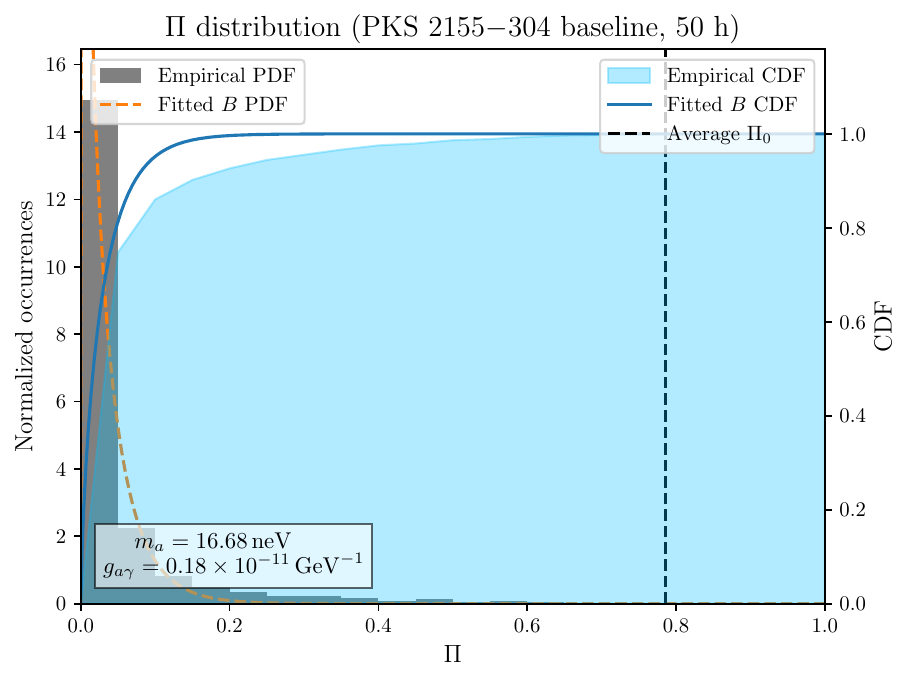}
    \caption{Example of a highly skewed $\Pi$ distribution obtained for a ML classifier close to maximal accuracy}
    \label{fig:example-ts-distribution-ml-toogood}
\end{figure}

In cases where the classifier accuracy is maximal, i.e. for large $\gag$ values, the resulting $\Pi$ distribution is largely peaked around very small values, as discussed in Section \ref{sec:ml-limits} (see e.g. Figure \ref{fig:example-ts-distribution-ml-toogood}). In such extreme occurrences, the fit to the Beta distribution is typically not as good (resulting in small $p$-values from the KS test), along with being very sensitive to the numerical precision of the data. Moreover, the average value $\Pi_0$ obtained from the ALP-less datasets is usually close to 1, resulting in a confidence level for excluding the ALP hypothesis of essentially 100\%, and consequently in a divergent $z$-score; see eqs. \eqref{eq:cl} and \eqref{eq:z-score}. 
A possible explanation of this behavior is found in the binary nature of the classifier algorithm, whose result is determined by averaging the predictions of several decision trees. When $\gag$ is small, the size of Poissonian fluctuations in the datasets is comparable with the ALP-related spectral features. In classifying the datasets, a decision tree applies splits on the features (in this case, photon counts in each bin) based on thresholds set during the learning process. Poissonian fluctuations may then cause the photon counts in certain bins to lie above or below these decision thresholds, leading different datasets to follow different splits along the tree and consequently forming a continuous distribution of $\Pi$ values. On the other hand, in regions of the ALP parameter space where spectral features are large, the difference between different datasets due only to Poissonian fluctuations may be too small compared to the decision thresholds in the trees. This leads almost all datasets to follow the same splits in each classification tree, ultimately resulting in very similar predictions. As a consequence, when computing the $\Pi$ statistic for 2\,000 test datasets at high $\gag$, a highly peaked distribution is found, poorly fit by a smooth shape such as the $\beta$ PDF.

In order to determine whether our main results hold under these limitations, we recomputed the exclusion confidence levels for each point in the ALP parameter space simply as the fraction of the 2\,000 simulated $\Pi(m_a, \gag)$ values smaller than $\Pi_0(m_a,\gag)$. This approach is in fact equivalent to eq. \eqref{eq:cl}, but it does not require to assume a specific distribution for $\Pi$. By doing so, we have explicitly checked that the 95\% CL ($2\sigma$) sensitivity region on the ALP parameter space remain unchanged. On the other hand, without the smoothing provided by a fitting function, the $z$-score diverges to infinity much more quickly as $\gag$ increases. We conclude that higher-confidence limits (e.g. $3\sigma$, $5\sigma$) obtained with the classifier-based method should be regarded as less reliable and more dependent on the quality of the fit.

%
\bibliographystyle{unsrt}
\bibliography{biblio}

@article{pecceiquinn1977,
  title = {{$\mathrm{CP}$ Conservation in the Presence of Pseudoparticles}},
  author = {Peccei, R. D. and Quinn, H. R.},
  journal = {Phys. Rev. Lett.},
  volume = {38},
  issue = {25},
  pages = {1440--1443},
  numpages = {0},
  year = {1977},
  month = {Jun},
  publisher = {American Physical Society},
  doi = {10.1103/PhysRevLett.38.1440},
  url = {https://link.aps.org/doi/10.1103/PhysRevLett.38.1440}
}

@article{weinberg1978,
  title = {{A New Light Boson?}},
  author = {Weinberg, S.},
  journal = {Phys. Rev. Lett.},
  volume = {40},
  issue = {4},
  pages = {223--226},
  numpages = {0},
  year = {1978},
  month = {Jan},
  publisher = {American Physical Society},
  doi = {10.1103/PhysRevLett.40.223},
  url = {https://link.aps.org/doi/10.1103/PhysRevLett.40.223}
}

@article{wilczek1978,
  title = {{Problem of Strong {$P$} and $T$ Invariance in the Presence of Instantons}},
  author = {Wilczek, F.},
  journal = {Phys. Rev. Lett.},
  volume = {40},
  issue = {5},
  pages = {279--282},
  numpages = {0},
  year = {1978},
  month = {Jan},
  publisher = {American Physical Society},
  doi = {10.1103/PhysRevLett.40.279},
  url = {https://link.aps.org/doi/10.1103/PhysRevLett.40.279}
}

@article{donnelly1978,
  title = {Do axions exist?},
  author = {Donnelly, T. W. and Freedman, S. J. and Lytel, R. S. and Peccei, R. D. and Schwartz, M.},
  journal = {Phys. Rev. D},
  volume = {18},
  issue = {5},
  pages = {1607--1620},
  numpages = {0},
  year = {1978},
  month = {Sep},
  publisher = {American Physical Society},
  doi = {10.1103/PhysRevD.18.1607},
  url = {https://link.aps.org/doi/10.1103/PhysRevD.18.1607}
}

@article{zehnder1981,
title = {{Axion search in a monochromatic $\gamma$-transition: A new lower limit for the axion mass}},
journal = {Phys. Lett. B},
volume = {104},
number = {6},
pages = {494-498},
year = {1981},
issn = {0370-2693},
doi = {https://doi.org/10.1016/0370-2693(81)90522-0},
url = {https://www.sciencedirect.com/science/article/pii/0370269381905220},
author = {A. Zehnder},
}

@article{dine1981,
title = {{A simple solution to the strong CP problem with a harmless axion}},
journal = {Phys. Lett. B},
volume = {104},
number = {3},
pages = {199-202},
year = {1981},
issn = {0370-2693},
doi = {https://doi.org/10.1016/0370-2693(81)90590-6},
url = {https://www.sciencedirect.com/science/article/pii/0370269381905906},
author = {M. Dine and W. Fischler and M. Srednicki},
}

@article{zhitnitsky1980,
    author = "Zhitnitsky, A. R.",
    title = "{On Possible Suppression of the Axion Hadron Interactions. (In Russian)}",
    journal = "Sov. J. Nucl. Phys.",
    volume = "31",
    pages = "260",
    year = "1980"
}

@article{kim1979,
  title = {{Weak-Interaction Singlet and Strong $\mathrm{CP}$ Invariance}},
  author = {Kim, J. E.},
  journal = {Phys. Rev. Lett.},
  volume = {43},
  issue = {2},
  pages = {103--107},
  numpages = {0},
  year = {1979},
  month = {Jul},
  publisher = {American Physical Society},
  doi = {10.1103/PhysRevLett.43.103},
  url = {https://link.aps.org/doi/10.1103/PhysRevLett.43.103}
}

@article{shifman1980,
title = {{Can confinement ensure natural CP invariance of strong interactions?}},
journal = {Nucl. Phys. B},
volume = {166},
number = {3},
pages = {493-506},
year = {1980},
issn = {0550-3213},
doi = {https://doi.org/10.1016/0550-3213(80)90209-6},
url = {https://www.sciencedirect.com/science/article/pii/0550321380902096},
author = {M. A. Shifman and A. I. Vainshtein and V. I. Zakharov},
}

@article{irastorza2018,
title = {New experimental approaches in the search for axion-like particles},
journal = {Prog. Part. Nucl. Phys.},
volume = {102},
pages = {89-159},
year = {2018},
issn = {0146-6410},
doi = {https://doi.org/10.1016/j.ppnp.2018.05.003},
url = {https://www.sciencedirect.com/science/article/pii/S014664101830036X},
author = {I. G. Irastorza and J. Redondo},
}

@article{jaeckel2010,
   author = {Jaeckel, J. and Ringwald, A.},
   title = "The Low-Energy Frontier of Particle Physics", 
   journal= "Annu. Rev. Nucl. Part. Sci.",
   year = "2010",
   volume = "60",
   number = "Volume 60, 2010",
   pages = "405-437",
   doi = "https://doi.org/10.1146/annurev.nucl.012809.104433",
   url = "https://www.annualreviews.org/content/journals/10.1146/annurev.nucl.012809.104433",
   publisher = "Annual Reviews",
   issn = "1545-4134",
   type = "Journal Article",
  }

@article{ringwald2012,
title = {{Exploring the role of axions and other WISPs in the dark universe}},
journal = {Phys. Dark Universe},
volume = {1},
number = {1},
pages = {116-135},
year = {2012},
note = {Next Decade in Dark Matter and Dark Energy},
issn = {2212-6864},
doi = {https://doi.org/10.1016/j.dark.2012.10.008},
url = {https://www.sciencedirect.com/science/article/pii/S221268641200012X},
author = {A. Ringwald},
}

@article{sikivie1983,
  title = {{Experimental Tests of the "Invisible" Axion}},
  author = {Sikivie, P.},
  journal = {Phys. Rev. Lett.},
  volume = {51},
  issue = {16},
  pages = {1415--1417},
  numpages = {0},
  year = {1983},
  month = {Oct},
  publisher = {American Physical Society},
  doi = {10.1103/PhysRevLett.51.1415},
  url = {https://link.aps.org/doi/10.1103/PhysRevLett.51.1415}
}

@article{raffelt1988,
  title = {Mixing of the photon with low-mass particles},
  author = {Raffelt, G. and Stodolsky, L.},
  journal = {Phys. Rev. D},
  volume = {37},
  issue = {5},
  pages = {1237--1249},
  numpages = {0},
  year = {1988},
  month = {Mar},
  publisher = {American Physical Society},
  doi = {10.1103/PhysRevD.37.1237},
  url = {https://link.aps.org/doi/10.1103/PhysRevD.37.1237}
}

@article{bahre2013,
doi = {10.1088/1748-0221/8/09/T09001},
url = {https://dx.doi.org/10.1088/1748-0221/8/09/T09001},
year = {2013},
month = {sep},
publisher = {},
volume = {8},
number = {09},
pages = {T09001},
author = {R. Bähre and B. Döbrich and J. Dreyling-Eschweiler and S. Ghazaryan and R. Hodajerdi and others},
title = {{Any light particle search II -- Technical Design Report}},
journal = {J. Instrum.},
}

@article{armengaud2019,
doi = {10.1088/1475-7516/2019/06/047},
url = {https://dx.doi.org/10.1088/1475-7516/2019/06/047},
year = {2019},
month = {jun},
publisher = {},
volume = {2019},
number = {06},
pages = {047},
author = {Armengaud, E. and Attié, D. and Basso, S. and Brun, P. and Bykovskiy, N. and others},
title = {{Physics potential of the International Axion Observatory (IAXO)}},
journal = {JCAP},
}

@article{du2018,
  title = {{Search for Invisible Axion Dark Matter with the Axion Dark Matter Experiment}},
  author = {Du, N. and Force, N. and Khatiwada, R. and Lentz, E. and Ottens, R. and others},
  collaboration = {ADMX},
  journal = {Phys. Rev. Lett.},
  volume = {120},
  issue = {15},
  pages = {151301},
  numpages = {5},
  year = {2018},
  month = {Apr},
  publisher = {American Physical Society},
  doi = {10.1103/PhysRevLett.120.151301},
  url = {https://link.aps.org/doi/10.1103/PhysRevLett.120.151301}
}

@article{brun2019,
   title={{A new experimental approach to probe QCD axion dark matter in the mass range above ${40}\,{\upmu}\mathrm{{eV}}$}},
   volume={79},
   ISSN={1434-6052},
   url={http://dx.doi.org/10.1140/epjc/s10052-019-6683-x},
   DOI={10.1140/epjc/s10052-019-6683-x},
   number={3},
   journal={Eur. Phys. J. C},
   publisher={Springer Science and Business Media LLC},
   author={Brun, P. and Caldwell, A. and Chevalier, L. and Dvali, G. and Freire, P. and others},
  collaboration = {MADMAX},
   year={2019},
   month=mar 
}

@article{deangelis2008,
title = {Axion-like particles, cosmic magnetic fields and gamma-ray astrophysics},
journal = {Phys. Lett. B},
volume = {659},
number = {5},
pages = {847-855},
year = {2008},
issn = {0370-2693},
doi = {https://doi.org/10.1016/j.physletb.2007.12.012},
url = {https://www.sciencedirect.com/science/article/pii/S0370269307015171},
author = {A. {De Angelis} and O. Mansutti and M. Roncadelli}
}

@article{horns2012,
  title = {{Hardening of TeV gamma spectrum of active galactic nuclei in galaxy clusters by conversions of photons into axionlike particles}},
  author = {Horns, D. and Maccione, L. and Meyer, M. and Mirizzi, A. and Montanino, D. and Roncadelli, M.},
  journal = {Phys. Rev. D},
  volume = {86},
  issue = {7},
  pages = {075024},
  numpages = {12},
  year = {2012},
  month = {Oct},
  publisher = {American Physical Society},
  doi = {10.1103/PhysRevD.86.075024},
  url = {https://link.aps.org/doi/10.1103/PhysRevD.86.075024}
}

@Inbook{mirizzi2008,
author="Mirizzi, A.
and Raffelt, G. G.
and Serpico, P. D.",
editor="Kuster, M.
and Raffelt, G.
and Beltr{\'a}n, B.",
title="Photon-Axion Conversion in Intergalactic Magnetic Fields and Cosmological Consequences",
bookTitle="Axions: Theory, Cosmology, and Experimental Searches",
year="2008",
publisher="Springer Berlin Heidelberg",
address="Berlin, Heidelberg",
pages="115--134",
isbn="978-3-540-73518-2",
doi="10.1007/978-3-540-73518-2_7",
url="https://doi.org/10.1007/978-3-540-73518-2_7"
}

@Article{galanti2022,
AUTHOR = {Galanti, G. and Roncadelli, M.},
TITLE = {{Axion-like Particles Implications for High-Energy Astrophysics}},
JOURNAL = {Universe},
VOLUME = {8},
YEAR = {2022},
NUMBER = {5},
ARTICLE-NUMBER = {253},
URL = {https://www.mdpi.com/2218-1997/8/5/253},
ISSN = {2218-1997},
DOI = {10.3390/universe8050253}
}

@Article{batkovic2021,
AUTHOR = {Batković, I. and De Angelis, A. and Doro, M. and Manganaro, M.},
TITLE = {{Axion-like Particle Searches with IACTs}},
JOURNAL = {Universe},
VOLUME = {7},
YEAR = {2021},
NUMBER = {6},
ARTICLE-NUMBER = {185},
URL = {https://www.mdpi.com/2218-1997/7/6/185},
ISSN = {2218-1997},
DOI = {10.3390/universe7060185}
}

@article{breitwheeler1934,
  title = {Collision of Two Light Quanta},
  author = {Breit, G. and Wheeler, J. A.},
  journal = {Phys. Rev.},
  volume = {46},
  issue = {12},
  pages = {1087--1091},
  numpages = {0},
  year = {1934},
  month = {Dec},
  publisher = {American Physical Society},
  doi = {10.1103/PhysRev.46.1087},
  url = {https://link.aps.org/doi/10.1103/PhysRev.46.1087}
}

@article{deangelis2013,
    author = {De Angelis, A. and Galanti, G. and Roncadelli, M.},
    title = {Transparency of the Universe to gamma-rays},
    journal = {Mon. Not. R. Astron. Soc.},
    volume = {432},
    number = {4},
    pages = {3245-3249},
    year = {2013},
    month = {05},
    issn = {0035-8711},
    doi = {10.1093/mnras/stt684},
    url = {https://doi.org/10.1093/mnras/stt684},
    eprint = {https://academic.oup.com/mnras/article-pdf/432/4/3245/18603541/stt684.pdf},
}

@article{padovani2017,
   title={Active galactic nuclei: what’s in a name?},
   volume={25},
   ISSN={1432-0754},
   url={http://dx.doi.org/10.1007/s00159-017-0102-9},
   DOI={10.1007/s00159-017-0102-9},
   number={1},
   journal={Astron. Astrophys. Rev.},
   publisher={Springer Science and Business Media LLC},
   author={Padovani, P. and Alexander, D. M. and Assef, R. J. and De Marco, B. and Giommi, P. and others},
   year={2017},
   month=aug }

@article{meyer2014a,
doi = {10.1088/1475-7516/2014/09/003},
url = {https://dx.doi.org/10.1088/1475-7516/2014/09/003},
year = {2014},
month = {sep},
publisher = {},
volume = {2014},
number = {09},
pages = {003},
author = {M. Meyer and D. Montanino and J. Conrad},
title = {On detecting oscillations of gamma rays into axion-like particles  in turbulent and coherent magnetic fields},
journal = {JCAP},
}

@article{meyer2014b,
doi = {10.1088/1475-7516/2014/12/016},
url = {https://dx.doi.org/10.1088/1475-7516/2014/12/016},
year = {2014},
month = {dec},
publisher = {},
volume = {2014},
number = {12},
pages = {016},
author = {M. Meyer and J. Conrad},
title = {{Sensitivity of the Cherenkov Telescope Array to the detection of axion-like particles at high gamma-ray opacities}},
journal = {JCAP},
}

@article{ajello2016,
  title = {{Search for Spectral Irregularities due to Photon--Axionlike-Particle Oscillations with the Fermi Large Area Telescope}},
  author = {Ajello, M. and Albert, A. and Anderson, B. and Baldini, L. and Barbiellini, G. and others},
  collaboration = {The Fermi-LAT},
  journal = {Phys. Rev. Lett.},
  volume = {116},
  issue = {16},
  pages = {161101},
  numpages = {7},
  year = {2016},
  month = {Apr},
  publisher = {American Physical Society},
  doi = {10.1103/PhysRevLett.116.161101},
  url = {https://link.aps.org/doi/10.1103/PhysRevLett.116.161101}
}

@article{abramowski2013,
  title = {{Constraints on axionlike particles with H.E.S.S. from the irregularity of the PKS $2155\ensuremath{-}304$ energy spectrum}},
  author = {Abramowski, A. and Acero, F. and Aharonian, F. and {Ait Benkhali}, F. and Akhperjanian, A. G. and others},
  collaboration = {H.E.S.S.},
  journal = {Phys. Rev. D},
  volume = {88},
  issue = {10},
  pages = {102003},
  numpages = {12},
  year = {2013},
  month = {Nov},
  publisher = {American Physical Society},
  doi = {10.1103/PhysRevD.88.102003},
  url = {https://link.aps.org/doi/10.1103/PhysRevD.88.102003}
}

@article{abdalla2021,
doi = {10.1088/1475-7516/2021/02/048},
url = {https://dx.doi.org/10.1088/1475-7516/2021/02/048},
year = {2021},
month = {feb},
publisher = {},
volume = {2021},
number = {02},
pages = {048},
author = {H. Abdalla and H. Abe and F. Acero and A. Acharyya and R. Adam and others},
title = {{Sensitivity of the Cherenkov Telescope Array for probing cosmology and fundamental physics with gamma-ray propagation}},
journal = {JCAP},
}

@article{davies2023,
  title = {{Constraints on axionlike particles from a combined analysis of three flaring $Fermi$ flat-spectrum radio quasars}},
  author = {Davies, J. and Meyer, M. and Cotter, G.},
  journal = {Phys. Rev. D},
  volume = {107},
  issue = {8},
  pages = {083027},
  numpages = {17},
  year = {2023},
  month = {Apr},
  publisher = {American Physical Society},
  doi = {10.1103/PhysRevD.107.083027},
  url = {https://link.aps.org/doi/10.1103/PhysRevD.107.083027}
}

@article{abe2024,
title = {{Constraints on axion-like particles with the Perseus Galaxy Cluster with MAGIC}},
journal = {Phys. Dark Universe},
volume = {44},
pages = {101425},
year = {2024},
issn = {2212-6864},
doi = {https://doi.org/10.1016/j.dark.2024.101425},
url = {https://www.sciencedirect.com/science/article/pii/S2212686424000074},
author = {H. Abe and S. Abe and J. Abhir and V. A. Acciari and I. Agudo and others},
}

@article{jacobsen2023,
doi = {10.1088/1475-7516/2023/10/009},
url = {https://dx.doi.org/10.1088/1475-7516/2023/10/009},
year = {2023},
month = {oct},
publisher = {IOP Publishing},
volume = {2023},
number = {10},
pages = {009},
author = {Jacobsen, S. and Linden, T. and Freese, K.},
title = {{Constraining axion-like particles with HAWC observations of TeV blazars}},
journal = {JCAP},
}

@article{adams2021,
title = {{Detection of the Crab Nebula with the 9.7 m prototype Schwarzschild-Couder telescope}},
journal = {Astropart. Phys.},
volume = {128},
pages = {102562},
year = {2021},
issn = {0927-6505},
doi = {https://doi.org/10.1016/j.astropartphys.2021.102562},
url = {https://www.sciencedirect.com/science/article/pii/S0927650521000062},
author = {C. B. Adams and R. Alfaro and G. Ambrosi and M. Ambrosio and C. Aramo and others},
}

@article{acharya2013,
title = {{Introducing the CTA concept}},
journal = {Astropart. Phys.},
volume = {43},
pages = {3-18},
year = {2013},
note = {Seeing the High-Energy Universe with the Cherenkov Telescope Array - The Science Explored with the CTA},
issn = {0927-6505},
doi = {https://doi.org/10.1016/j.astropartphys.2013.01.007},
url = {https://www.sciencedirect.com/science/article/pii/S0927650513000169},
author = {B. S. Acharya and M. Actis and T. Aghajani and G. Agnetta and J. Aguilar and others},
}

@book{ctascience2018,
   author = {B. S. Acharya and I. Agudo and I. Al Samarai and R. Alfaro and J. Alfaro and others},
   title={{Science with the Cherenkov Telescope Array}},
   ISBN={9789813270091},
   url={http://dx.doi.org/10.1142/10986},
   DOI={10.1142/10986},
   publisher={World Scientific},
   year={2018},
   month=feb 
}

@article{hooper2007,
  title = {Detecting Axionlike Particles with Gamma Ray Telescopes},
  author = {Hooper, D. and Serpico, P. D.},
  journal = {Phys. Rev. Lett.},
  volume = {99},
  issue = {23},
  pages = {231102},
  numpages = {4},
  year = {2007},
  month = {Dec},
  publisher = {American Physical Society},
  doi = {10.1103/PhysRevLett.99.231102},
  url = {https://link.aps.org/doi/10.1103/PhysRevLett.99.231102}
}

@article{franceschini2008,
	author = {Franceschini, A. and Rodighiero, G. and Vaccari, M.},
	title = {Extragalactic optical-infrared background radiation, its time evolution and the cosmic photon-photon opacity},
	DOI= "10.1051/0004-6361:200809691",
	url= "https://doi.org/10.1051/0004-6361:200809691",
	journal = {A\&A},
	year = 2008,
	volume = 487,
	number = 3,
	pages = "837-852",
}

@article{dominguez2011,
    author = {Domínguez, A. and Primack, J. R. and Rosario, D. J. and Prada, F. and Gilmore, R. C. and others},
    title = {{Extragalactic background light inferred from AEGIS galaxy-SED-type fractions}},
    journal = {Mon. Not. R. Astron. Soc.},
    volume = {410},
    number = {4},
    pages = {2556-2578},
    year = {2011},
    month = {01},
    issn = {0035-8711},
    doi = {10.1111/j.1365-2966.2010.17631.x},
    url = {https://doi.org/10.1111/j.1365-2966.2010.17631.x},
    eprint = {https://academic.oup.com/mnras/article-pdf/410/4/2556/6295256/mnras0410-2556.pdf},
}

@article{deangelis2007,
  title = {Evidence for a new light spin-zero boson from cosmological gamma-ray propagation?},
  author = {De Angelis, A. and Roncadelli, M. and Mansutti, O.},
  journal = {Phys. Rev. D},
  volume = {76},
  issue = {12},
  pages = {121301},
  numpages = {4},
  year = {2007},
  month = {Dec},
  publisher = {American Physical Society},
  doi = {10.1103/PhysRevD.76.121301},
  url = {https://link.aps.org/doi/10.1103/PhysRevD.76.121301}
}

@article{mirizzi2007,
  title = {{Signatures of axionlike particles in the spectra of TeV gamma-ray sources}},
  author = {Mirizzi, A. and Raffelt, G. G. and Serpico, P. D.},
  journal = {Phys. Rev. D},
  volume = {76},
  issue = {2},
  pages = {023001},
  numpages = {7},
  year = {2007},
  month = {Jul},
  publisher = {American Physical Society},
  doi = {10.1103/PhysRevD.76.023001},
  url = {https://link.aps.org/doi/10.1103/PhysRevD.76.023001}
}

@article{deangelis2011,
  title = {Relevance of axionlike particles for very-high-energy astrophysics},
  author = {De Angelis, A. and Galanti, G. and Roncadelli, M.},
  journal = {Phys. Rev. D},
  volume = {84},
  issue = {10},
  pages = {105030},
  numpages = {37},
  year = {2011},
  month = {Nov},
  publisher = {American Physical Society},
  doi = {10.1103/PhysRevD.84.105030},
  url = {https://link.aps.org/doi/10.1103/PhysRevD.84.105030}
}

@Article{bottcher2019,
AUTHOR = {B\"ottcher, M.},
TITLE = {{Progress in Multi-Wavelength and Multi-Messenger Observations of Blazars and Theoretical Challenges}},
JOURNAL = {Galaxies},
VOLUME = {7},
YEAR = {2019},
NUMBER = {1},
ARTICLE-NUMBER = {20},
URL = {https://www.mdpi.com/2075-4434/7/1/20},
ISSN = {2075-4434},
DOI = {10.3390/galaxies7010020}
}

@article{4fgl-dr3,
doi = {10.3847/1538-4365/ac6751},
url = {https://dx.doi.org/10.3847/1538-4365/ac6751},
year = {2022},
month = {jun},
publisher = {The American Astronomical Society},
volume = {260},
number = {2},
pages = {53},
author = {Abdollahi, S. and Acero, F. and Baldini, L. and Ballet, J. and Bastieri, D. and others},
title = {{Incremental Fermi Large Area Telescope Fourth Source Catalog}},
journal = {ApJS},
}

@misc{4fgl-dr4,
      title={{Fermi Large Area Telescope Fourth Source Catalog Data Release 4 (4FGL-DR4)}}, 
      author={J. Ballet and P. Bruel and T. H. Burnett and B. Lott and {The Fermi-LAT collaboration}},
      year={2024},
      eprint={2307.12546},
      archivePrefix={arXiv},
      primaryClass={astro-ph.HE},
      url={https://arxiv.org/abs/2307.12546}, 
}

@article{albert2007,
doi = {10.1086/521382},
url = {https://dx.doi.org/10.1086/521382},
year = {2007},
month = {nov},
publisher = {},
volume = {669},
number = {2},
pages = {862},
author = {Albert, J. and Aliu, E. and Anderhub, H. and Antoranz, P. and Armada, A. and others},
title = {{Variable Very High Energy $\gamma$-Ray Emission from Markarian 501}},
journal = {Astrophys. J.},
}

@article{aharonian2009,
	author = {Aharonian, F. and Akhperjanian, A. G. and Anton, G. and Barres de Almeida, U. and Bazer-Bachi, A. R. and others},
	title = {{Simultaneous multiwavelength observations of the second exceptional
 $\gamma$-ray flare of PKS 2155$–$304 in July 2006}},
	DOI= "10.1051/0004-6361/200912128",
	url= "https://doi.org/10.1051/0004-6361/200912128",
	journal = {Astron. Astrophys.},
	year = 2009,
	volume = 502,
	number = 3,
	pages = "749-770",
}

@article{gilmore2012,
    author = {Gilmore, R. C. and Somerville, R. S. and Primack, J. R. and Domínguez, A.},
    title = {{Semi-analytic modelling of the extragalactic background light and consequences for extragalactic gamma-ray spectra}},
    journal = {Mon. Not. R. Astron. Soc.},
    volume = {422},
    number = {4},
    pages = {3189-3207},
    year = {2012},
    month = {05},
    issn = {0035-8711},
    doi = {10.1111/j.1365-2966.2012.20841.x},
    url = {https://doi.org/10.1111/j.1365-2966.2012.20841.x},
    eprint = {https://academic.oup.com/mnras/article-pdf/422/4/3189/18601416/mnras0422-3189.pdf},
}

@article{saldanalopez2021,
    author = {Saldana-Lopez, A. and Domínguez, A. and Pérez-González, P. G. and Finke, J. and Ajello, M. and others},
    title = {{An observational determination of the evolving extragalactic background light from the multiwavelength HST/CANDELS survey in the Fermi and CTA era}},
    journal = {Mon. Not. R. Astron. Soc.},
    volume = {507},
    number = {4},
    pages = {5144-5160},
    year = {2021},
    month = {08},
    issn = {0035-8711},
    doi = {10.1093/mnras/stab2393},
    url = {https://doi.org/10.1093/mnras/stab2393},
    eprint = {https://academic.oup.com/mnras/article-pdf/507/4/5144/40391540/stab2393.pdf},
}

@article{finke2010,
doi = {10.1088/0004-637X/712/1/238},
url = {https://dx.doi.org/10.1088/0004-637X/712/1/238},
year = {2010},
month = {feb},
publisher = {The American Astronomical Society},
volume = {712},
number = {1},
pages = {238},
author = {Finke, J. D. and Razzaque, S. and Dermer, C. D.},
title = {Modeling the extragalactic background light from stars and dust},
journal = {Astrophys. J.},
}

@article{meyer2021,
  author = "Meyer, M.  and  Davies, J.  and  Kuhlmann, J.",
  title = "{gammaALPs: An open-source python package for computing photon-axion-like-particle oscillations in astrophysical environments}",
  doi = "10.22323/1.395.0557",
  journal = "PoS",
  year = 2021,
  volume = "ICRC2021",
  pages = "557"
}

@article{davies2021,
  title = {Relevance of jet magnetic field structure for blazar axionlike particle searches},
  author = {Davies, J. and Meyer, M. and Cotter, G.},
  journal = {Phys. Rev. D},
  volume = {103},
  issue = {2},
  pages = {023008},
  numpages = {16},
  year = {2021},
  month = {Jan},
  publisher = {American Physical Society},
  doi = {10.1103/PhysRevD.103.023008},
  url = {https://link.aps.org/doi/10.1103/PhysRevD.103.023008}
}

@article{potter2013a,
    author = {Potter, W. J. and Cotter, G.},
    title = {{Synchrotron and inverse-Compton emission from blazar jets – I. A uniform conical jet model}},
    journal = {Mon. Not. R. Astron. Soc.},
    volume = {423},
    number = {1},
    pages = {756-765},
    year = {2012},
    month = {05},
    issn = {0035-8711},
    doi = {10.1111/j.1365-2966.2012.20918.x},
    url = {https://doi.org/10.1111/j.1365-2966.2012.20918.x},
    eprint = {https://academic.oup.com/mnras/article-pdf/423/1/756/18611634/mnras0423-0756.pdf},
}

@article{potter2013b,
    author = {Potter, W. J. and Cotter, G.},
    title = {{Synchrotron and inverse-Compton emission from blazar jets – II. An accelerating jet model with a geometry set by observations of M87}},
    journal = {Mon. Not. R. Astron. Soc.},
    volume = {429},
    number = {2},
    pages = {1189-1205},
    year = {2012},
    month = {12},
    issn = {0035-8711},
    doi = {10.1093/mnras/sts407},
    url = {https://doi.org/10.1093/mnras/sts407},
    eprint = {https://academic.oup.com/mnras/article-pdf/429/2/1189/18454914/sts407.pdf},
}

@article{potter2013c,
    author = {Potter, W. J. and Cotter, G.},
    title = "{Synchrotron and inverse-Compton emission from blazar jets – III. Compton-dominant blazars}",
    journal = {Mon. Not. R. Astron. Soc.},
    volume = {431},
    number = {2},
    pages = {1840-1852},
    year = {2013},
    month = {03},
    issn = {0035-8711},
    doi = {10.1093/mnras/stt300},
    url = {https://doi.org/10.1093/mnras/stt300},
    eprint = {https://academic.oup.com/mnras/article-pdf/431/2/1840/4282729/stt300.pdf},
}

@article{potter2013d,
    author = {Potter, W. J. and Cotter, G.},
    title = "{Synchrotron and inverse-Compton emission from blazar jets – IV. BL Lac type blazars and the physical basis for the blazar sequence}",
    journal = {Mon. Not. R. Astron. Soc.},
    volume = {436},
    number = {1},
    pages = {304-314},
    year = {2013},
    month = {10},
    issn = {0035-8711},
    doi = {10.1093/mnras/stt1569},
    url = {https://doi.org/10.1093/mnras/stt1569},
    eprint = {https://academic.oup.com/mnras/article-pdf/436/1/304/18496736/stt1569.pdf},
}

@article{potter2015,
    author = {Potter, W. J. and Cotter, G.},
    title = {{New constraints on the structure and dynamics of black hole jets}},
    journal = {Mon. Not. R. Astron. Soc.},
    volume = {453},
    number = {4},
    pages = {4070-4088},
    year = {2015},
    month = {09},
    issn = {0035-8711},
    doi = {10.1093/mnras/stv1657},
    url = {https://doi.org/10.1093/mnras/stv1657},
    eprint = {https://academic.oup.com/mnras/article-pdf/453/4/4070/8033656/stv1657.pdf},
}

@article{tavecchio2010,
    author = {Tavecchio, F. and Ghisellini, G. and Ghirlanda, G. and Foschini, L. and Maraschi, L.},
    title = {{TeV BL Lac objects at the dawn of the Fermi era}},
    journal = {Mon. Not. R. Astron. Soc.},
    volume = {401},
    number = {3},
    pages = {1570-1586},
    year = {2010},
    month = {01},
    issn = {0035-8711},
    doi = {10.1111/j.1365-2966.2009.15784.x},
    url = {https://doi.org/10.1111/j.1365-2966.2009.15784.x},
    eprint = {https://academic.oup.com/mnras/article-pdf/401/3/1570/3811367/mnras0401-1570.pdf},
}

@article{abdo2009,
doi = {10.1088/0067-0049/183/1/46},
url = {https://dx.doi.org/10.1088/0067-0049/183/1/46},
year = {2009},
month = {jun},
publisher = {The American Astronomical Society},
volume = {183},
number = {1},
pages = {46},
author = {Abdo, A. A. and Ackermann, M. and Ajello, M. and Atwood, W. B. and Axelsson, M. and others},
title = {{Fermi/Large Area Telescope Bright Gamma-Ray Source List}},
journal = {ApJS},
}

@article{abdo2010,
doi = {10.1088/0004-637X/716/1/30},
url = {https://dx.doi.org/10.1088/0004-637X/716/1/30},
year = {2010},
month = {may},
publisher = {The American Astronomical Society},
volume = {716},
number = {1},
pages = {30},
author = {Abdo, A. A. and Ackermann, M. and Agudo, I. and Ajello, M. and Aller, H. D. and others},
title = {{The Spectral Energy Distribution of Fermi Bright Blazars}},
journal = {Astrophys. J.},
}

@article{jansson2012,
doi = {10.1088/0004-637X/757/1/14},
url = {https://dx.doi.org/10.1088/0004-637X/757/1/14},
year = {2012},
month = {aug},
publisher = {The American Astronomical Society},
volume = {757},
number = {1},
pages = {14},
author = {Jansson, R. and Farrar, G. R.},
title = {{A new model of the Galactic magnetic field}},
journal = {Astrophys. J.},
}

@article{ade2016,
	author = {Ade, P. A. R. and Aghanim, N. and Arnaud, M. and Arroja, F. and Ashdown, M. and others},
	title = {{Planck 2015 results - XIX. Constraints on primordial magnetic fields}},
    collaboration = {Planck},
	DOI= "10.1051/0004-6361/201525821",
	url= "https://doi.org/10.1051/0004-6361/201525821",
	journal = {Astron. Astrophys.},
	year = 2016,
	volume = 594,
	pages = "A19",
}

@article{gammapy2023,
 author = {{Donath}, A. and {Terrier}, R. and {Remy}, Q. and {Sinha}, A. and {Nigro}, C. and
 others},
 title = {{Gammapy: A Python package for gamma-ray astronomy}},
 DOI= "10.1051/0004-6361/202346488",
 url= "https://doi.org/10.1051/0004-6361/202346488",
 journal = {Astron. Astrophys.},
 year = 2023,
 volume = 678,
 pages = "A157",
 }

@online{gammapy-1.0.1,
  author       = {Acero, F. and
                  Aguasca-Cabot, A. and
                  Buchner, J. and
                  Carreto Fidalgo, D. and
                  Chen, A. and others},
  title        = {{Gammapy: Python toolbox for gamma-ray astronomy}},
  month        = mar,
  year         = 2023,
  publisher    = {Zenodo},
  version      = {v1.0.1},
  doi          = {10.5281/zenodo.7734804},
  url          = {https://doi.org/10.5281/zenodo.7734804},
}

@online{ctao_irfs_prod5,
  author       = {{Cherenkov Telescope Array Observatory and
                  Cherenkov Telescope Array Consortium}},
  title        = {{CTAO Instrument Response Functions - prod5 version
                   v0.1
                  }},
  month        = sep,
  year         = 2021,
  publisher    = {Zenodo},
  version      = {v0.1},
  doi          = {10.5281/zenodo.5499840},
  url          = {https://doi.org/10.5281/zenodo.5499840},
}

@article{cowan2011,
   title={Asymptotic formulae for likelihood-based tests of new physics},
   volume={71},
   ISSN={1434-6052},
   url={http://dx.doi.org/10.1140/epjc/s10052-011-1554-0},
   DOI={10.1140/epjc/s10052-011-1554-0},
   number={2},
   journal={Eur. Phys. J. C},
   publisher={Springer Science and Business Media LLC},
   author={Cowan, G. and Cranmer, K. and Gross, E. and Vitells, O.},
   year={2011},
   month=feb }

@article{acharyya2021,
doi = {10.1088/1475-7516/2021/01/057},
url = {https://dx.doi.org/10.1088/1475-7516/2021/01/057},
year = {2021},
month = {jan},
publisher = {},
volume = {2021},
number = {01},
pages = {057},
author = {Acharyya, A. and Adam, R. and Adams, C. and Agudo, I. and Aguirre-Santaella, A. and others},
title = {{Sensitivity of the Cherenkov Telescope Array to a dark matter signal from the Galactic centre}},
journal = {JCAP},
}

@article{abe2024b,
doi = {10.1088/1475-7516/2024/07/047},
url = {https://dx.doi.org/10.1088/1475-7516/2024/07/047},
year = {2024},
month = {jul},
publisher = {IOP Publishing},
volume = {2024},
number = {07},
pages = {047},
author = {Abe, S. and Abhir, J. and Abhishek, A. and Acero, F. and Acharyya, A. and others},
title = {{Dark matter line searches with the Cherenkov Telescope Array}},
journal = {JCAP},
}

@article{acharyya2023,
    author = {Acharyya, A. and Adam, R. and Aguasca-Cabot, A. and Agudo, I. and Aguirre-Santaella, A. and others},
    title = {{Sensitivity of the Cherenkov Telescope Array to TeV photon emission from the Large Magellanic Cloud}},
    journal = {Mon. Not. R. Astron. Soc.},
    volume = {523},
    number = {4},
    pages = {5353-5387},
    year = {2023},
    month = {05},
    issn = {0035-8711},
    doi = {10.1093/mnras/stad1576},
    url = {https://doi.org/10.1093/mnras/stad1576},
    eprint = {https://academic.oup.com/mnras/article-pdf/523/4/5353/50733482/stad1576.pdf},
}

@article{astropy2013,
Adsurl = {http://adsabs.harvard.edu/abs/2013A%26A...558A..33A},
Archiveprefix = {arXiv},
Author = {{Astropy Collaboration} and others},
Doi = {10.1051/0004-6361/201322068},
Eid = {A33},
Eprint = {1307.6212},
Journal = {Astron. Astrophys.},
Month = oct,
Pages = {A33},
Primaryclass = {astro-ph.IM},
Title = {{Astropy: A community Python package for astronomy}},
Volume = 558,
Year = 2013}

@ARTICLE{astropy2018,
       author = {{Astropy Collaboration} and others},
        title = "{The Astropy Project: Building an Open-science Project and Status of the v2.0 Core Package}",
      journal = {Astrophys. J.},
         year = 2018,
        month = sep,
       volume = {156},
       number = {3},
          eid = {123},
        pages = {123},
          doi = {10.3847/1538-3881/aabc4f},
archivePrefix = {arXiv},
       eprint = {1801.02634},
 primaryClass = {astro-ph.IM},
       adsurl = {https://ui.adsabs.harvard.edu/abs/2018AJ....156..123A}
}

@ARTICLE{astropy2022,
       author = {{Astropy Collaboration} and others},
        title = "{The Astropy Project: Sustaining and Growing a Community-oriented Open-source Project and the Latest Major Release (v5.0) of the Core Package}",
      journal = {Astrophys. J.},
         year = 2022,
        month = aug,
       volume = {935},
       number = {2},
          eid = {167},
        pages = {167},
          doi = {10.3847/1538-4357/ac7c74},
archivePrefix = {arXiv},
       eprint = {2206.14220},
 primaryClass = {astro-ph.IM},
       adsurl = {https://ui.adsabs.harvard.edu/abs/2022ApJ...935..167A}
}

@ARTICLE{wakely2008,
author = {{Wakely}, S.~P. and {Horan}, D.},
title = "{TeVCat: An online catalog for Very High Energy Gamma-Ray Astronomy}",
journal = {International Cosmic Ray Conference},
year = 2008,
volume = 3,
pages = {1341-1344},
adsurl = {http://adsabs.harvard.edu/abs/2008ICRC....3.1341W}
}

@article{bassan2010,
doi = {10.1088/1475-7516/2010/05/010},
url = {https://dx.doi.org/10.1088/1475-7516/2010/05/010},
year = {2010},
month = {may},
publisher = {},
volume = {2010},
number = {05},
pages = {010},
author = {N. Bassan and A. Mirizzi and M. Roncadelli},
title = {Axion-like particle effects on the polarization of cosmic high-energy gamma sources},
journal = {JCAP},
}

@article{wilks1938,
author = {S. S. Wilks},
title = {{The Large-Sample Distribution of the Likelihood Ratio for Testing Composite Hypotheses}},
volume = {9},
journal = {Ann. Math. Statist.},
number = {1},
publisher = {Institute of Mathematical Statistics},
pages = {60 -- 62},
year = {1938},
doi = {10.1214/aoms/1177732360},
URL = {https://doi.org/10.1214/aoms/1177732360}
}

@ARTICLE{virtanen2020,
  author  = {Virtanen, P. and Gommers, R. and Oliphant, T. E. and
            Haberland, M. and Reddy, T. and others},
  title   = {{{SciPy} 1.0: Fundamental Algorithms for Scientific
            Computing in Python}},
  journal = {Nat. Methods},
  year    = {2020},
  volume  = {17},
  pages   = {261--272},
  adsurl  = {https://rdcu.be/b08Wh},
  doi     = {10.1038/s41592-019-0686-2},
}

@article{day2020,
   title={Accelerating the search for axion-like particles with machine learning},
   volume={2020},
   ISSN={1475-7516},
   url={http://dx.doi.org/10.1088/1475-7516/2020/03/046},
   DOI={10.1088/1475-7516/2020/03/046},
   number={03},
   journal={JCAP},
   publisher={IOP Publishing},
   author={Day, F. and Krippendorf, S.},
   year={2020},
   month=mar, pages={046–046} 
}

@book{raschka2017,  
address = {Birmingham, UK},  
author = {Raschka, S. and Mirjalili, V.},  
edition = {2},  
isbn = {978-1787125933},  
publisher = {Packt Publishing},  
title = {{Python Machine Learning, 2nd Ed.}},  
year = {2017}  
}

@article{ohm2009,
title = {{$\gamma$/hadron separation in very-high-energy $\gamma$-ray astronomy using a multivariate analysis method}},
journal = {Astropart. Phys.},
volume = {31},
number = {5},
pages = {383-391},
year = {2009},
issn = {0927-6505},
doi = {https://doi.org/10.1016/j.astropartphys.2009.04.001},
url = {https://www.sciencedirect.com/science/article/pii/S0927650509000589},
author = {S. Ohm and C. {van Eldik} and K. Egberts},
}

@article{albert2008,
title = {{Implementation of the Random Forest method for the Imaging Atmospheric Cherenkov Telescope MAGIC}},
journal = {Nucl. Instrum. Meth.},
volume = {A588},
number = {3},
pages = {424-432},
year = {2008},
issn = {0168-9002},
doi = {https://doi.org/10.1016/j.nima.2007.11.068},
url = {https://www.sciencedirect.com/science/article/pii/S0168900207024059},
author = {J. Albert and E. Aliu and H. Anderhub and P. Antoranz and A. Armada and others},
}

@article{bock2004,
title = {{Methods for multidimensional event classification: a case study using images from a Cherenkov gamma-ray telescope}},
journal = {Nucl. Instrum. Meth.},
volume = {A516},
number = {2},
pages = {511-528},
year = {2004},
issn = {0168-9002},
doi = {https://doi.org/10.1016/j.nima.2003.08.157},
url = {https://www.sciencedirect.com/science/article/pii/S0168900203025051},
author = {R. K. Bock and A. Chilingarian and M. Gaug and F. Hakl and T. Hengstebeck and others},
}

@INPROCEEDINGS{hillas1985,
       author = {{Hillas}, A.~M.},
        title = "{Cerenkov Light Images of EAS Produced by Primary Gamma Rays and by Nuclei}",
    booktitle = {19th International Cosmic Ray Conference (ICRC19), Volume 3},
         year = 1985,
       editor = {{Jones}, F.~C.},
       series = {International Cosmic Ray Conference},
       volume = {3},
        month = aug,
        pages = {445},
       adsurl = {https://ui.adsabs.harvard.edu/abs/1985ICRC....3..445H}
}

@article{pedregosa2012,
  title={{Scikit-learn: Machine Learning in {P}ython}},
  author={Pedregosa, F. and Varoquaux, G. and Gramfort, A. and Michel, V. and Thirion, B. and others},
  journal={J. Mach. Learn. Res.},
  volume={12},
  pages={2825--2830},
  year={2011}
}

@inproceedings{chen2016, 
   series={KDD ’16},
   title={{XGBoost: A Scalable Tree Boosting System}},
   url={http://dx.doi.org/10.1145/2939672.2939785},
   DOI={10.1145/2939672.2939785},
   booktitle={Proceedings of the 22nd ACM SIGKDD International Conference on Knowledge Discovery and Data Mining},
   publisher={ACM},
   author={Chen, T. and Guestrin, C.},
   year={2016},
   month=aug, pages={785–794},
   collection={KDD ’16} }

@article{diluzio2020,
title = {{The landscape of QCD axion models}},
journal = {Phys. Rep.},
volume = {870},
pages = {1-117},
year = {2020},
issn = {0370-1573},
doi = {https://doi.org/10.1016/j.physrep.2020.06.002},
url = {https://www.sciencedirect.com/science/article/pii/S0370157320302477},
author = {L. {Di Luzio} and M. Giannotti and E. Nardi and L. Visinelli},
}

@article{noordhuis2023,
  title = {Novel Constraints on Axions Produced in Pulsar Polar-Cap Cascades},
  author = {Noordhuis, D. and Prabhu, A. and Witte, S. J. and Chen, A. Y. and Cruz, F. and Weniger, C.},
  journal = {Phys. Rev. Lett.},
  volume = {131},
  issue = {11},
  pages = {111004},
  numpages = {7},
  year = {2023},
  month = {Sep},
  publisher = {American Physical Society},
  doi = {10.1103/PhysRevLett.131.111004},
  url = {https://link.aps.org/doi/10.1103/PhysRevLett.131.111004}
}

@article{dessert2022,
  title = {Upper limit on the axion-photon coupling from magnetic white dwarf polarization},
  author = {Dessert, C. and Dunsky, D. and Safdi, B. R.},
  journal = {Phys. Rev. D},
  volume = {105},
  issue = {10},
  pages = {103034},
  numpages = {22},
  year = {2022},
  month = {May},
  publisher = {American Physical Society},
  doi = {10.1103/PhysRevD.105.103034},
  url = {https://link.aps.org/doi/10.1103/PhysRevD.105.103034}
}

@article{manzari2024,
  title = {Supernova Axions Convert to Gamma Rays in Magnetic Fields of Progenitor Stars},
  author = {Manzari, C. A. and Park, Y. and Safdi, B. R. and Savoray, I.},
  journal = {Phys. Rev. Lett.},
  volume = {133},
  issue = {21},
  pages = {211002},
  numpages = {10},
  year = {2024},
  month = {Nov},
  publisher = {American Physical Society},
  doi = {10.1103/PhysRevLett.133.211002},
  url = {https://link.aps.org/doi/10.1103/PhysRevLett.133.211002}
}

@misc{AxionLimits,
  author       = {C. O'Hare},
  title        = {cajohare/AxionLimits: AxionLimits},
  month        = jul,
  year         = 2020,
  publisher    = {Zenodo},
  version      = {v1.0},
  doi          = {10.5281/zenodo.3932430},
  howpublished = {\url{https://cajohare.github.io/AxionLimits/}}
}

@software{gammapy-1.3,
  author       = {Acero, F. and
                  Aguasca-Cabot, A. and
                  Bernete, J. and
                  Biederbeck, N. and
                  Djuvsland, J. and others},
  title        = {Gammapy: Python toolbox for gamma-ray astronomy},
  month        = jan,
  year         = 2025,
  publisher    = {Zenodo},
  version      = {v1.3},
  doi          = {10.5281/zenodo.14760974},
  url          = {https://doi.org/10.5281/zenodo.14760974},
  swhid        = {swh:1:dir:9405c2435b92c1267179543aaf913e9a3fda6ced
                   ;origin=https://doi.org/10.5281/zenodo.4701488;vis
                   it=swh:1:snp:ee94f46fa10032e542469d3614f2699ea257e
                   827;anchor=swh:1:rel:707c1b3d2919e9f78dbb649ad584c
                   30eec8359d2;path=gammapy-gammapy-c8b5337
                  },
}

@article{kohri2017,
  title = {Axion-like particles and recent observations of the cosmic infrared background radiation},
  author = {Kohri, K. and Kodama, H.},
  journal = {Phys. Rev. D},
  volume = {96},
  issue = {5},
  pages = {051701},
  numpages = {6},
  year = {2017},
  month = {Sep},
  publisher = {American Physical Society},
  doi = {10.1103/PhysRevD.96.051701},
  url = {https://link.aps.org/doi/10.1103/PhysRevD.96.051701}
}

@article{zioutas1999,
title = {A decommissioned LHC model magnet as an axion telescope},
journal = {Nucl. Instrum. Methods Phys. Res. A},
volume = {425},
number = {3},
pages = {480-487},
year = {1999},
issn = {0168-9002},
doi = {https://doi.org/10.1016/S0168-9002(98)01442-9},
url = {https://www.sciencedirect.com/science/article/pii/S0168900298014429},
author = {K. Zioutas and C. E. Aalseth and D. Abriola and F. T. {Avignone III} and R. L. Brodzinski and others},
}

@article{altenmuller2024,
  title = {New Upper Limit on the Axion-Photon Coupling with an Extended CAST Run with a Xe-Based Micromegas Detector},
  author = {Altenm\"uller, K. and Anastassopoulos, V. and Arguedas-Cuendis, S. and Aune, S. and Baier, J. and others},
  collaboration = {CAST Collaboration},
  journal = {Phys. Rev. Lett.},
  volume = {133},
  issue = {22},
  pages = {221005},
  numpages = {8},
  year = {2024},
  month = {Nov},
  publisher = {American Physical Society},
  doi = {10.1103/PhysRevLett.133.221005},
  url = {https://link.aps.org/doi/10.1103/PhysRevLett.133.221005}
}

@Article{galanti2024,
AUTHOR = {Galanti, G.},
TITLE = {Axion-like Particle Effects on Photon Polarization in High-Energy Astrophysics},
JOURNAL = {Universe},
YEAR = {2024},
NUMBER = {8},
ARTICLE-NUMBER = {312},
URL = {https://www.mdpi.com/2218-1997/10/8/312},
ISSN = {2218-1997},
DOI = {10.3390/universe10080312}
}

@Article{cerruti2020,
AUTHOR = {Cerruti, M.},
TITLE = {Leptonic and Hadronic Radiative Processes in Supermassive-Black-Hole Jets},
JOURNAL = {Galaxies},
VOLUME = {8},
YEAR = {2020},
NUMBER = {4},
ARTICLE-NUMBER = {72},
URL = {https://www.mdpi.com/2075-4434/8/4/72},
ISSN = {2075-4434},
DOI = {10.3390/galaxies8040072}
}

@article{bottcher2013,
doi = {10.1088/0004-637X/768/1/54},
url = {https://doi.org/10.1088/0004-637X/768/1/54},
year = {2013},
month = {apr},
publisher = {The American Astronomical Society},
volume = {768},
number = {1},
pages = {54},
author = {B\"ottcher, M. and Reimer, A. and Sweeney, K. and Prakash, A.},
title = {Leptonic and hadronic modeling of {\it Fermi}-detected blazars},
journal = {Astrophys. J.},
}

@ARTICLE{li1983,
       author = {{Li}, T.-P. and {Ma}, Y.-Q.},
        title = "{Analysis methods for results in gamma-ray astronomy.}",
      journal = {Astrophys. J.},
         year = 1983,
        month = sep,
       volume = {272},
        pages = {317-324},
          doi = {10.1086/161295},
       adsurl = {https://ui.adsabs.harvard.edu/abs/1983ApJ...272..317L}
}

@article{loporchio2025,
doi = {10.1088/1742-6596/3053/1/012020},
url = {https://doi.org/10.1088/1742-6596/3053/1/012020},
year = {2025},
month = {jul},
publisher = {IOP Publishing},
volume = {3053},
number = {1},
pages = {012020},
author = {Loporchio, S.},
title = {CTA+: an Italian program to enhance the Southern Cherenkov Telescope Array Observatory},
journal = {Journal of Physics: Conference Series},
}

@Article{choudhury2024,
author={Choudhury, A.
and Mondal, A.
and Sarkar, S.},
title="{Searches for the BSM scenarios at the LHC using decision tree-based machine learning algorithms: a comparative study and review of random forest, AdaBoost, XGBoost and LightGBM frameworks}",
journal={Eur. Phys. J. Spec. Top.},
year={2024},
month={Nov},
day={01},
volume={233},
number={15},
pages={2425-2463},
issn={1951-6401},
doi={10.1140/epjs/s11734-024-01308-x},
url={https://doi.org/10.1140/epjs/s11734-024-01308-x}
}

@article{chao2019,
title = "{Study of Star/Galaxy Classification Based on the XGBoost Algorithm}",
journal = {Chin. Astron. Astrophys.},
volume = {43},
number = {4},
pages = {539-548},
year = {2019},
issn = {0275-1062},
doi = {https://doi.org/10.1016/j.chinastron.2019.11.005},
url = {https://www.sciencedirect.com/science/article/pii/S0275106219300815},
author = {L. Chao and Z. Wen-hui and L. Ji-ming},
}

@article{tolamatti2023,
    author = {Tolamatti, A. and Singh, K. K. and Yadav, K. K.},
    title = "{Classification of the Fermi-LAT blazar candidates of uncertain type using extreme gradient boosting}",
    journal = {Mon. Not. R. Astron. Soc.},
    volume = {523},
    number = {4},
    pages = {5341-5352},
    year = {2023},
    month = {06},
    issn = {0035-8711},
    doi = {10.1093/mnras/stad1826},
    url = {https://doi.org/10.1093/mnras/stad1826},
    eprint = {https://academic.oup.com/mnras/article-pdf/523/4/5341/50725024/stad1826.pdf},
}

@article{zhou2025,
doi = {10.1088/1475-7516/2025/08/084},
url = {https://doi.org/10.1088/1475-7516/2025/08/084},
year = {2025},
month = {aug},
publisher = {IOP Publishing},
volume = {2025},
number = {08},
pages = {084},
author = {Zhou, T. and Gao, L.-Q. and Bi, X.-J. and Yin, P.-F. and Lin, W.},
title = "{Exploring constraints on axion-like particles with the observations for blazar Mrk 501}",
journal = {JCAP},
}

@article{abe2025,
    author = {Abe, K. and Abe, S. and Abhishek, A. and Acero, F. and Aguasca-Cabot, A. and others},
    title = "{VHE $\gamma$-ray observations of bright BL Lacs with the Large-Sized Telescope prototype (LST-1) of the CTAO}",
    journal = {Mon. Not. R. Astron. Soc.},
    volume = {544},
    number = {1},
    pages = {669-686},
    year = {2025},
    month = {10},
    issn = {0035-8711},
    doi = {10.1093/mnras/staf1728},
    url = {https://doi.org/10.1093/mnras/staf1728},
    eprint = {https://academic.oup.com/mnras/article-pdf/544/1/669/64573083/staf1728.pdf},
}

@article{gao2025,
doi = {10.1088/1475-7516/2025/01/031},
url = {https://doi.org/10.1088/1475-7516/2025/01/031},
year = {2025},
month = {jan},
publisher = {IOP Publishing},
volume = {2025},
number = {01},
pages = {031},
author = {Gao, L.-Q. and Bi, X.-J. and Li, J. and Yin, P.-F.},
title = {Impact of parameters in the toroidal blazar jet magnetic field model on axion-like particle constraints},
journal = {JCAP},
}

\end{document}